%% file: root_v2.tex
\documentclass{article}

\usepackage{arxiv}

\usepackage{cite}
\usepackage{amsmath,amssymb,amsbsy,amsfonts}%,amsthm}
\usepackage{graphicx}
\usepackage{textcomp}
\usepackage{algorithm}
\usepackage{algpseudocode}
\usepackage{tabularx,booktabs}
\usepackage{adjustbox}
\usepackage{array,etoolbox}
\usepackage{dcolumn}
\usepackage{rotating}
\usepackage{pifont}
\usepackage{tikz}
\usepackage{xcolor}
\usepackage{comment}
\usepackage{hyperref}
\usepackage{cuted}

\usepackage{multirow}
\usepackage{array}
\usepackage{color, colortbl}

\newcolumntype{L}[1]{>{\raggedright\let\newline\\\arraybackslash\hspace{0pt}}m{#1}}
\newcolumntype{C}[1]{>{\centering\let\newline\\\arraybackslash\hspace{0pt}}m{#1}}
\newcolumntype{R}[1]{>{\raggedleft\let\newline\\\arraybackslash\hspace{0pt}}m{#1}}

\newcolumntype{Y}{>
{\centering\arraybackslash}X}

\def\vsp{\vspace*}

\newcommand{\eqq}{$\,=\,$}

\newcommand\sbullet[1][.5]{\mathbin{\vcenter{\hbox{\scalebox{#1}{$\bullet$}}}}}

\usepackage{pifont}
\newcommand{\xmark}{\ding{54}}
\newcommand\scross[1][0.5]{\mathbin{\vcenter{\hbox{\scalebox{#1}{\xmark}}}}}

\newcommand\sadd[1][0.7]{\mathbin{\vcenter{\hbox{\scalebox{#1}{$+$}}}}}
\newcommand\sstar[1][0.8]{\mathbin{\vcenter{\hbox{\scalebox{#1}{$*$}}}}}
\newcommand\stria[1][0.8]{\mathbin{\vcenter{\hbox{\scalebox{#1}{$\triangledown$}}}}}
\newcommand\striu[1][0.7]{\mathbin{\vcenter{\hbox{\scalebox{#1}{$\triangle$}}}}}
\newcommand\ssqur[1][0.7]{\mathbin{\vcenter{\hbox{\scalebox{#1}{$\square$}}}}}

\DeclareMathAlphabet\mathbfcal{OMS}{cmsy}{b}{n}

\newcommand{\ben}{\begin{eqnarray*}}
	\newcommand{\een}{\end{eqnarray*}}
\newcommand{\be}{\begin{eqnarray}}
\newcommand{\ee}{\end{eqnarray}}

\newcommand{\bv}{\mathbf{v}}
\newcommand{\bx}{\mathbf{x}}
\newcommand{\bhx}{\hat{\mathbf{x}}}

\newcommand{\bz}{\mathbf{z}}

\newcommand{\bw}{\mathbf{w}}

\newcommand{\bB}{\mathbf{B}}
\newcommand{\bQ}{\mathbf{Q}}
\newcommand{\bP}{\mathbf{P}}

\newcommand{\bbN}{\mathbb{N}}
\newcommand{\bbR}{\mathbb{R}}

\newcommand{\ccN}{\mathcal{N}}

\newcommand{\bet}{\boldsymbol{\eta}}
\newcommand{\bhet}{\hat{\boldsymbol{\eta}}}

\newcommand{\ka}{\kappa}

\newcommand{\LR}{\star}
\newcommand{\TR}{}

\usepackage{acro}
	\input{notation}

\renewcommand{\shorttitle}{Incorporating BTL into PF for Dual-Tracking System with Asymmetric Noise Intensities}

\begin{document}
%\title{Bayesian Transfer Learning with Particle Filter for Object Tracking under Asymmetric Noise Intensities}
\title{Incorporating Bayesian Transfer Learning into Particle Filter for Dual-Tracking System with Asymmetric Noise Intensities}
\author{Omar A. Alotaibi,  Brian L.\ Mark, and Mohammad Reza Fasihi
\thanks{A preliminary version of this work was presented in~\cite{Alotaibi-2025-TL_PF}. 
The authors are with the Department of Electrical and Computer Engineering, George Mason University, Fairfax, VA 22030 USA (e-mail: oalotaib@gmu.edu; bmark@gmu.edu; mfasihi4@gmu.edu).}
}
%\thanks{Author }}

% The paper headers
\markboth{IEEE TRANSACTIONS, VOL. XX, NO. XX, XXXX 2025}{Alotaibi and Mark:}

\maketitle

\begin{abstract}
Using Bayesian transfer learning, we develop a particle filter approach for tracking a nonlinear dynamical model in a dual-tracking system where intensities of measurement noise for both sensors are asymmetric.  The densities for Bayesian transfer learning are approximated with the sum of weighted particles to improve the tracking performance of the primary sensor, which experiences a higher noise intensity compared to the source sensor.  We present simulation results that validate the effectiveness of the proposed approach compared to an isolated particle filter and transfer learning applied to the unscented Kalman filter and the cubature Kalman filter. Furthermore, increasing the number of particles shows an improvement in the performance of transfer learning applied to the particle filter with a higher rate compared to the isolated particle filter. However, increasing the number of particles raises computational time per step. Moreover, the performance gain from incorporating Bayesian transfer learning is approximately linearly proportional to the absolute difference value between the noise intensities of the sensors in the dual-tracking system. 
\end{abstract}


\acresetall

\input{Intro}

\input{Dynamic_Systems}

\input{BTLF}

\input{PF_TL_Section}

\input{PF_Sim_v2}

\input{Conclusion}

\appendix
\input{Proof_TL_PF}

%\section*{References}
\bibliographystyle{IEEEtran}
\bibliography{IEEEabrv,library}

\end{document}

%% file: notation.tex
\DeclareAcronym{wf}{
	short = WF ,
	long  = Wiener Filter ,
}
\DeclareAcronym{fir}{
	short = FIR ,
	long  = Finite Impulse Response ,
}
\DeclareAcronym{wss}{
	short = WSS ,
	long  = Wide Sense Stationary ,
}
\DeclareAcronym{wna}{
	short = WNA ,
	long  = White Noise Acceleration ,
}
\DeclareAcronym{mmse}{
	short = MMSE ,
	long  = Minimum Mean Square Error ,
}
\DeclareAcronym{rmse}{
	short = RMSE ,
	long  = root-mean square error ,
}
\DeclareAcronym{mse}{
	short = MSE ,
	long  = mean square error ,
}
\DeclareAcronym{ssr}{
	short = SSR ,
	long  = state space representation ,
}
\DeclareAcronym{iid}{
	short = i.i.d. ,
	long  = independent and identically distributed ,
}
\DeclareAcronym{pdf}{
	short = PDF ,
	long  = probability density function ,
}
\DeclareAcronym{pdfs}{
	short = PDFs ,
	long  = probability density functions ,
}
\DeclareAcronym{pmf}{
	short = PMF ,
	long  = probability mass function ,
}
\DeclareAcronym{kf}{
	short = KF ,
	long  = Kalman filter ,
}
\DeclareAcronym{ekf}{
	short = EKF ,
	long  = extended Kalman filter ,
}
\DeclareAcronym{ukf}{
	short = UKF ,
	long  = unscented Kalman filter ,
}
\DeclareAcronym{ut}{
	short = UT ,
	long  = unscented transformation ,
}
\DeclareAcronym{pf}{
	short = PF ,
	long  = particle filter ,
}
\DeclareAcronym{sis}{
	short = SIS ,
	long  = Sequential Importance Sampling ,
}
\DeclareAcronym{sir}{
	short = SIR ,
	long  = Sequential Importance Resampling ,
}
\DeclareAcronym{sis-pf}{
	short = SIS-PF ,
	long  = sequential importance sampling particle filter ,
}
\DeclareAcronym{sir-pf}{
	short = SIR-PF ,
	long  = sequential importance resampling particle filter ,
}
\DeclareAcronym{mc}{
	short = MC ,
	long  = Monte Carlo ,
}
\DeclareAcronym{mcmc}{
	short = MCMC ,
	long  = Markov chain Monte Carlo ,
}
\DeclareAcronym{smc}{
	short = SMC ,
	long  = Sequential Monte Carlo ,
}
\DeclareAcronym{2d}{
	short = 2D ,
	long  = two dimensional ,
}
\DeclareAcronym{3d}{
	short = 3D ,
	long  = three dimensional ,
}
\DeclareAcronym{snr}{
	short = SNR ,
	long  = signal-to-noise ratio ,
}
\DeclareAcronym{apf}{
	short = APF ,
	long  = auxiliary particle filter ,
}
\DeclareAcronym{awgn}{
	short = AWGN ,
	long  = additive white Gaussian noise ,
}
\DeclareAcronym{tbd}{
	short = TBD ,
	long  = track-before-detect ,
}
\DeclareAcronym{pf-tbd}{
	short = PF-TBD ,
	long  = particle filter based track-before-detect ,
}
\DeclareAcronym{af}{
	short = AF ,
	long  = ambiguity function ,
}
\DeclareAcronym{fov}{
	short = FOV ,
	long  = field of view ,
}
\DeclareAcronym{rf}{
	short = RF ,
	long  = radio frequency ,
}
\DeclareAcronym{eo}{
	short = EO ,
	long  = electro-optical ,
}
\DeclareAcronym{eef}{
	short = EEF ,
	long  = exponentially embedded family ,
}
\DeclareAcronym{kl}{
	short = KL ,
	long  = Kullback-Leibler ,
}
\DeclareAcronym{mh}{
	short = M-H ,
	long  = Metropolis-Hastings ,
}
\DeclareAcronym{rfs}{
	short = RFS ,
	long  = random finite set ,
}
\DeclareAcronym{phdf}{
	short = PHDF ,
	long  = probability hypothesis density filter ,
}
\DeclareAcronym{phdf-pf}{
	short = PHDF-PF ,
	long  = particle implementation of the probability hypothesis density filter,
}
\DeclareAcronym{phd}{
	short = PHD ,
	long  = probability hypothesis density,
}
\DeclareAcronym{pgfl}{
	short = PGFL ,
	long  = probability generating functional ,
}
\DeclareAcronym{gmm}{
	short = GMM ,
	long  = Gaussian mixture model ,
}
\DeclareAcronym{memberf}{
	short = MeMBerF ,
	long  = multi-object multi-Bernoulli filter ,
}
\DeclareAcronym{cbmemberf}{
	short = CBMeMBerF ,
	long  = cardinality balanced multi-object multi-Bernoulli filter ,
}
\DeclareAcronym{cbmemberf-pf}{
	short = CBMeMBerF-PF ,
	long  = particle implementation of the cardinality balanced multi-object multi-Bernoulli filter ,
}
\DeclareAcronym{omat}{
	short = OMAT ,
	long  = optimal mass transfer ,
}
\DeclareAcronym{ospa}{
	short = OSPA ,
	long  = optimal subpattern assignment ,
}
\DeclareAcronym{em}{
	short = EM ,
	long  = Expectation Maximization ,
}
\DeclareAcronym{dp}{
	short = DP ,
	long  = Dirichlet Process ,
}
\DeclareAcronym{dpm}{
	short = DPM,
	long  = Dirichlet process mixture,
}
\DeclareAcronym{niwd}{
	short = NIWD ,
	long  = Normal-inverse-Wishart distribution ,
}

\DeclareAcronym{hdp}{
	short = HDP ,
	long  = Hierarchical Dirichlet Process ,
}

\DeclareAcronym{crp}{
	short = CRP ,
	long  = Chinese restaurant process ,
}

\DeclareAcronym{tl}{
	short = TL ,
	long  = transfer learning ,
}

\DeclareAcronym{tl-ukf}{
	short = TL-UKF ,
	long  = transfer learning unscented Kalman filter ,
}

\DeclareAcronym{ckf}{
	short = CKF ,
	long  = cubature Kalman filter ,
}

\DeclareAcronym{tl-ckf}{
	short = TL-CKF ,
	long  = transfer learning cubature Kalman filter ,
}

\DeclareAcronym{mvf}{
	short = MVF ,
	long  = measurement vector fusion ,
}

\DeclareAcronym{btl}{
	short = BTL ,
	long  = Bayesian transfer learning ,
}

\DeclareAcronym{btlf}{
	short = BTLF ,
	long  = Bayesian transfer learning filter ,
}

\DeclareAcronym{fpd}{
	short = FPD ,
	long  = fully probabilistic design ,
}

\DeclareAcronym{tl-pf}{
	short = TL-PF ,
	long  = transfer learning particle filter ,
}

\DeclareAcronym{dkf}{
	short = DKF ,
	long  = distributed Kalman filtering ,
}

\DeclareAcronym{dtnet}{
	short = DTNet ,
	long  = Detection Tracking Network ,
}

\DeclareAcronym{v2v}{
	short = V2V ,
	long  = vehicle-to-vehicle ,
}

\DeclareAcronym{lhs}{
	short = L.H.S ,
	long  = left-hand side ,
}

\DeclareAcronym{rhs}{
	short = R.H.S ,
	long  = right-hand side ,
}

%% file: Intro.tex
\section{Introduction}

Tracking systems have recently received significant attention due to the growing demand for higher accuracy in various applications such as object tracking, surveillance, navigation, self-driving vehicles, and computer vision.  In object tracking applications, observed data from multiple sensors is often combined using various data fusion techniques to improve overall tracking accuracy~\cite{Bar-Shalom:1986,Roecker:1988,Hashemipour:1988}. In scenarios of a dual separated tracking system, where two tracking systems track the same area at the same time, the two tracking systems are typically assumed to experience similar measurement noise intensity. However, this assumption may not always hold true as each tracking system can encounter various environmental conditions in practice.

%\IEEEPARstart{M}{ulti-sensor} systems have recently granted significant studies attention due to the growing demand of achieving higher accuracy performances across a variety of applications from multiple fields such as object tracking, surveillance, navigation, and computer vision. In object tracking applications, observed data from multiple sensors is often combined using various data fusion techniques to improve the overall tracking accuracy~\cite{Bar-Shalom:1986,Roecker:1988,Hashemipour:1988}. Traditionally, sensors within a multi-sensor system are typically assumed to experience similar measurement noise intensity. However, the validity of this assumption may not be consistently accurate, as sensors can encounter diverse environmental conditions in practice. 

We consider a scenario of dual tracking systems for tracking a single object, where the fields of view of the two separated tracking systems overlap. One of the dual-tracking systems is referred to as the primary tracking filter, while the other is the source tracking filter. Each of the tracking systems employs a particle filter to track the common object along a nonlinear trajectory.  We shall assume that the primary tracking filter's \ac{fov} encounters higher noise intensity or clutter in its measurements compared to that of the source tracking filter. Our aim is to improve the tracking performance of the primary tracking filter by leveraging knowledge from the source tracking filter using \ac{btl}.

%We consider a dual-sensor system for tracking a single object, where the fields of view of the two sensors overlap. One of the sensors is referred to as the primary sensor, while the other is the source sensor.  Each of the sensors employs a particle filter to track the common object along a nonlinear trajectory.  We shall assume that the primary sensor's \ac{fov} encounters higher noise intensity or clutter in its measurements compared to that of the source sensor. Our aim is to improve the tracking performance of the primary sensor by leveraging knowledge from the source sensor using \ac{btl}. 

\subsection{Motivation}
{\em Transfer learning} or {\em knowledge transfer}, a machine learning technique, has recently been introduced into Bayesian models involving multiple tracking systems~\cite{Pan-2009-survey_TL}. This approach is referred to as \ac{btl}~\cite{Karbalayghareh-2018-optimal_TL,Wu-2023-bayesian_TL,Zhou-2020-BTL,Zhao-2025-BTL}. Transfer learning allows knowledge gained from a source domain to improve the performance of a target domain that has the challenge of performing its own task accurately. \ac{btl} has been adopted to model Gaussian process regression with multiple tasks to effectively address the interactions between the source and target tasks~\cite{Karbalayghareh-2018-TL_regression,Papevz-2022-BTL_Gaussian_Process}. Furthermore, online object tracking has been modeled using \ac{btl} to allow the transfer of visual priors into the tracking process framework to improve performance under complex conditions~\cite{Wang-2012-TL_tracking_image}. 

A real-world scenario for online transfer learning occurs when multiple operators have tracking systems covering the same area of interest. Such systems often have different configurations, preventing them from processing raw observations from other operators. In this case, common variables, which are predicted observations, can be widely adapted worldwide to be transferred between these operators to enhance each other tracking performance. For instance, ships in a complex maritime environment leverage transferred information for the purpose of enhancing their detection and tracking performance using the \ac{dtnet}~\cite{He-2025-Maritime}. Furthermore, autonomous vehicles of different manufacturers can cooperate through \ac{v2v} communication in order to improve tracking performance by recursively sharing useful information~\cite{Chen-2020-V2V,Wang-2025-V2V}.

\ac{btl} has been applied to a pair of Kalman filters with asymmetric noise intensities for tracking an object following a {\em linear} motion model~\cite{Foley-2017_fully,Papevz-2019-robust,Papevz-2021-hierarchical,Skalsky-2025-BTL_Kalman,Huang-2025-TL-Klaman}. The incorporation of \ac{btl} with a Bayesian filter, referred to as a \ac{btlf}, aims to enhance the estimation performance of the primary tracking filter, whose \ac{fov} is affected by severe conditions, by leveraging gained knowledge from the source tracking filter. In the \ac{btlf} framework, the primary tracking filter has no access to any realization of the source tracking system except the observation predictor of the source tracking filter according to~\cite{Foley-2017_fully}. \ac{btlf} has been previously applied in~\cite{Alotaibi-2024-TL_UKF} to track a {\em nonlinear} motion model using a {\em local} approximation approach, where the posterior density follows the form of the prior density.  In this approach, the predicted observation parameters, i.e., the mean and covariance, obtained from Bayesian filtering using an unscented Kalman filter, were transferred from the source tracking filter to the primary tracking filter.  The \ac{btl} approach is different  
from \ac{mvf}~\cite{Willner-1976-MVF} and distributed estimation filtering methods such as \ac{dkf}~\cite{Olfati-2005_DKF,Olfati-2007_DKF,Talebi-2018_DKF}, in which the source tracking system directly transfers its estimated states or measurements to the primary tracking system. The drawbacks of \ac{mvf} and \ac{dkf} relative to the \ac{btlf} approach are as follows: 
(i) higher communication overhead in transferring cluttered observation data, 
(ii) one time-step delay in the transferred information, and
(iii) from a security or privacy perspective, the raw observation data may contain information that the source may not which to disclose.

\subsection{Contribution}
To the best of our knowledge, there have been no previous applications of {\em global} approaches, whereby no explicit assumption is made about the form of the posterior density, in the context of \ac{btlf}. A popular global approach for nonlinear filtering is the \ac{pf} with importance resampling~\cite{Gordon-1993-Introduced_PF, Gustafsson-2002-particle}. Our main contribution is to approximate \ac{btlf} via a \ac{pf} to track a nonlinear dynamical motion model in a dual-tracking system under asymmetric measurement noise intensities. Our simulation results show that the estimation accuracy performance of the proposed \ac{tl-pf} scheme, which incorporates \ac{btlf} with \ac{pf}, is significantly superior relative to the performance of an isolated \ac{pf} under the same conditions. Notably, the \ac{tl-pf} performance increases at a faster rate than the isolated \ac{pf} as the number of particles increases, which illustrates the higher sensitivity of \ac{tl-pf}. The main benefit of the proposed \ac{tl-pf} approach is that it outperforms not only the isolated \ac{pf}, but also the integration of \ac{ukf}~\cite{JulieU-2000-UKF} and \ac{ckf}~\cite{Arasaratnam-2009-CKF} with \ac{btlf}, which we refer to as \ac{tl-ukf} and \ac{tl-ckf} respectively, as recently published in~\cite{Alotaibi-2024_TL_CKF_arXiv,Alotaibi-2025-Multi_UKF}. This demonstrates the advantages of integrating \ac{pf} with \ac{btlf} to achieve higher accuracy for the challenges of nonlinear motion tracking in dual-tracking systems.

\subsection{Organization and Notation}

The remainder of this article is organized as follows. Section~\ref{Sec:Dynamic_Systems} briefly presents the dynamical nonlinear motion model and formulates the problem of tracking an object using a dual-tracking system with a nonlinear model under asymmetric measurement noise intensities. In Section~\ref{Sec:BTL}, the \ac{btlf} framework is applied to a single object tracking in a dual-tracking system. Section~\ref{Sec:TL_PF} develops the \ac{tl-pf} scheme and provides procedural steps of the proposed algorithm for the source and primary tracking filters. The numerical results of linear and nonlinear scenarios that demonstrate the effectiveness of the proposed scheme compared to the related tracking filters are presented in Section~\ref{Sec:TL_PF_Sim}. In addition to numerical results, Section~\ref{Sec:TL_PF_Sim} provides the computational time for incorporating \ac{btl} framework to tracking filters in comparison to isolated filters. The article outcomes along with suggested future work are concluded in Section~\ref{Sec:conclusion}.

{\em Notation throughout this article:} Uppercase and lowercase letters represent scalars, bold lowercase letters denote vectors, and bold uppercase letters are matrices, $(\cdot)^{T}$ is the transpose operation of a matrix, $(\cdot)^{-1}$ represents
the inverse operation of a matrix, and $\text{diag}[\cdot]$ denotes the diagonal of a matrix. $\bbN$, $\bbN_{0}$, $\bbR$, and $\bbR^{+}$ denote the sets of natural, natural with zero, real, and positive real numbers, respectively.

%% file: Dynamic_Systems.tex
\section{Dynamic System Model for Object Tracking} \label{Sec:Dynamic_Systems}

\subsection{Nonlinear Motion Model}

Consider a discrete-time dynamical nonlinear system~\cite{Ho-1964-Bayesian} described in a state-space representation by the following equations:
\begin{align}
\bx_{k} &= f\left( \bx_{k-1}\right) + \bv_{k-1}  , \label{Pro_form_State} \\
\bz_{k} &= h\left( \bx_{k}\right) + \bw_k  , \label{Pro_form_meas} 
\end{align}
where $k\in \bbN_{0}$ denotes the time step of the discrete-time dynamical system. The state of the system
at time~$k$ is represented by $\bx_{k} \in \bbR^{n_\bx}$ and
$f: \bbR^{n_\bx} \rightarrow \bbR^{n_\bx}$ is a nonlinear state transition function. The measurement of the system observed at time~$k$ is denoted by a vector $\bz_k \in\bbR^{n_\bz}$, which is determined by a nonlinear function $h: \bbR^{n_\bx} \rightarrow \bbR^{n_\bz}$ of the state vector. The vectors $\bv_{k-1} \in \bbR^{n_\bx}$ and $\bw_k \in \bbR^{n_\bz}$ define the process and measurement noises characterizing the uncertainty in the dynamical system. The state $\bx_{k}$ and measurement $\bz_k$ vectors for a particular object tracking scenario are defined
in Section~\ref{Sec:TL_PF_Sim}.

\subsection{Dual-tracking System with Asymmetric Noise Intensities} \label{Sec:Problem_form}

\begin{figure}
	\centering
	\includegraphics[trim={0.85cm 0cm 0.5cm 0cm},clip, scale=0.5]%[scale=0.35]
        {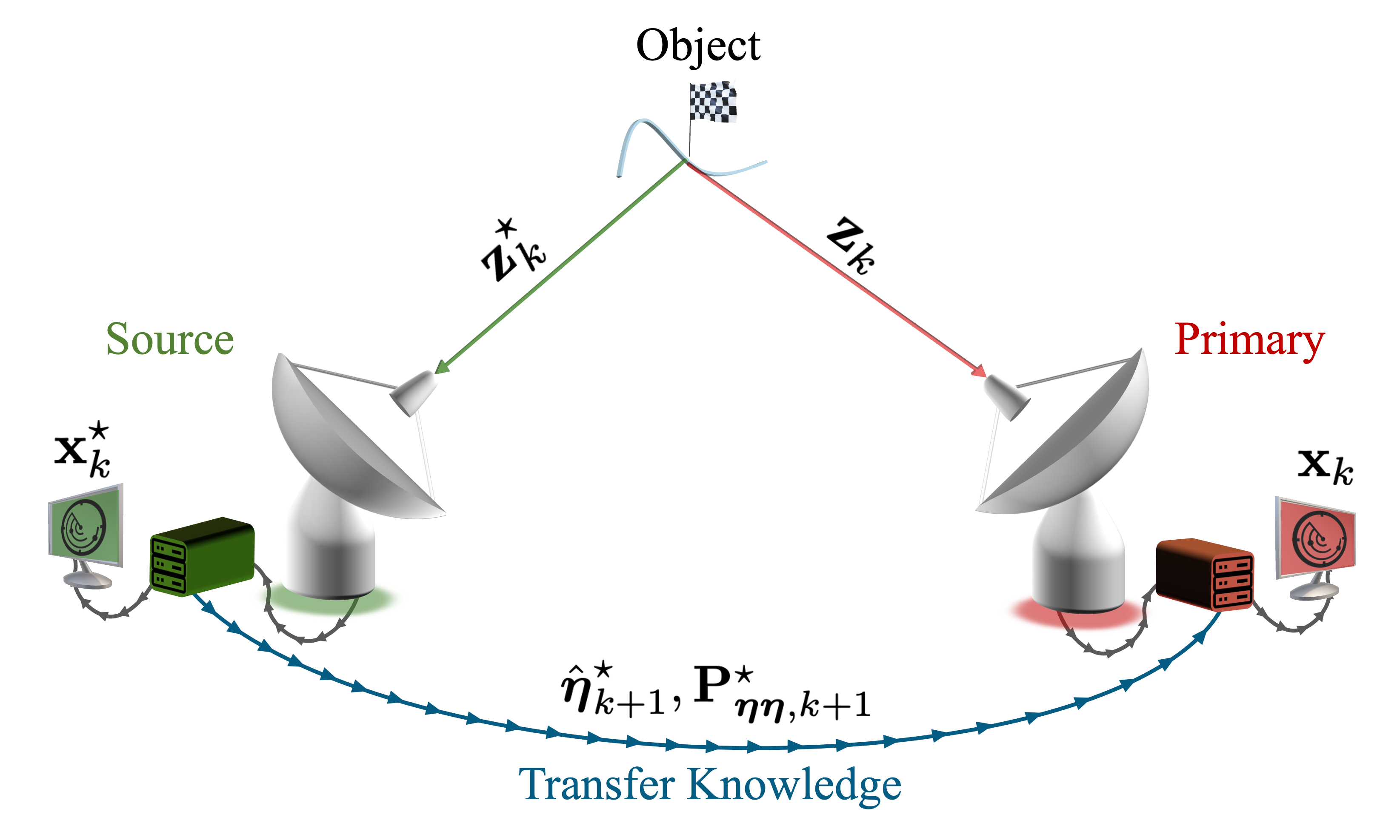}
	\caption{Visualization of a dual-tracking system with asymmetric noise intensities, illustrating the integration of transfer learning.}
	\label{fig:Problem_Form_3D}
\end{figure}

We consider a dual-tracking system tracking a single object following the nonlinear motion model provided in \eqref{Pro_form_State}. The process noise of both sensors is assumed to be \ac{iid} zero-mean Gaussian vector, expressed as $\bv_{k-1}  \stackrel{\rm iid}{\sim} \ccN (\mathbf{0}, \bQ_{\bv})$, with the same corresponding covariance $\bQ_{\bv} \in \bbR^{n_{\bx} \times n_{\bx}}$ for the source and the primary tracking filters. The two tracking systems track the desired object under the same type of measurement noise, which is assumed to be Gaussian. However, the environmental and surrounding conditions according to the location of each tracking system and its \ac{fov} are not necessarily the same. This variation in conditions leads to dissimilar measurement noise intensity that each tracking system experiences. For instance, two self-driving vehicles driving close to each other in the same area where each may be built by a different manufacturer. As shown in Fig.~\ref{fig:Problem_Form_3D}, the source tracking system measurements are denoted by $\bz^{\LR}_{k}$ with superscript $\LR$, while the primary tracking system measurements are denoted by $\bz_{k}$. 

The measurement model represented in \eqref{Pro_form_meas} will be separated for the source and primary tracking systems, respectively, as 
\begin{align}
\bz^{\LR}_{k} &= h\left( \bx_{k}\right) + \bw^{\LR}_k , \label{Pro_form_meas_LR} \\
\bz_{k} &= h\left( \bx_{k}\right) + \bw_k , \label{Pro_form_meas_TR}
\end{align}
where the measurement noises of the source and primary tracking systems are assumed to be \ac{iid} zero-mean Gaussian, i.e., 
\begin{align}
  \bw^{\LR}_{k} \stackrel{\rm iid}{\sim} \ccN (\mathbf{0}, \bQ^{\LR}_{\bw}) ~~~\mbox{and}~~~ \bw_{k} \stackrel{\rm iid}{\sim} \ccN (\mathbf{0}, \bQ_{\bw}) , \label{eq:noise_Gussian}
\end{align}
respectively. The noise covariances of the measurement model are given by $\bQ^{\LR}_{\bw}=I^{\LR}_{\bw} \bB_\bw$ and $\bQ_{\bw}=I_{\bw} \bB_\bw$, where $\bB_\bw \in \bbR^{n_{\bz} \times n_{\bz}}$ is a common diagonal matrix. The noise intensities of the source, $I^\LR_\bw \in \bbR^+$, and primary, $I_\bw \in \bbR^+$, tracking systems are assumed to be asymmetric, i.e., $I_{\bw} > I^{\LR}_{\bw}$, (see Section~\ref{Sec:TL_PF_Sim} for more definitions and details). The higher level of noise intensity at the primary sensor negatively impacts the reliability of its measurements and its tracking performance. Even though both tracking systems track the same object, their state estimations differ from each other due to the estimation process of each tracking filter conditioning on different observed measurement, with various noise intensity. This yields two estimated states where each tracking filter has its own state estimation. The source tracking filter estimation is denoted by $\bx^{\LR}_k$, while the primary tracking filter has an estimated state as $\bx_k$

%where $I^\LR_\bw \in \bbR^+$ and $I_\bw \in \bbR^+$ are the noise intensities of the source and primary sensors, respectively.  We assume that the noise intensities are asymmetric, i.e., $I_{\bw} > I^{\LR}_{\bw}$. The higher level of noise intensity at the primary sensor negatively impacts the reliability of its measurements and its tracking performance. the noise intensities $I_\bw , I^\LR_\bw \in \bbR^+$ and the common matrix $\bB_\bw \in \bbR^{n_{\bz} \times n_{\bz}}$. We assume that the noise intensities are asymmetric, i.e., $I_{\bw} > I^{\LR}_{\bw}$. The higher level of noise intensity at the primary sensor negatively impacts the reliability of its measurements and its tracking performance. 

%% file: BTLF.tex
\section{Bayesian Transfer Learning Filter Approach}
\label{Sec:BTL}

We briefly review the \ac{btlf} framework, which was applied in~\cite{Alotaibi-2024-TL_UKF,Alotaibi-2024_TL_CKF_arXiv} in
the context of a dual-tracking system based on the unscented Kalman filter. This framework builds on the concept of shared knowledge introduced in~\cite{Foley-2017_fully,Papevz-2022-BTL_Gaussian_Process} as refereed to \ac{fpd}, where the primary tracking filter is able to access the probabilistic parameters by transferring the predicted observation parameters, specifically mean $\bhet^{\LR}_{k+1}$ and covariance $\bP^{\LR}_{\bet\bet,k+1}$, from the source tracking filter.

\subsection{Source Tracking Filter}
In the source tracking filter, an unknown object state, denoted as $\bx^{\LR}_{k}$, of the current time step $k$ and a predicted observation $\bet^{\LR}_{k+1}$ of the next time step $k+1$ are estimated given a set of observed measurements up to time step $k$, defined as $\bz^{\LR}_{1:k}=\left\lbrace \bz^{\LR}_{1}, \bz^{\LR}_{2}, \ldots, \bz^{\LR}_{k}\right\rbrace $. The overall posterior density of the object state and predicted observation $p(\bx^{\LR}_{k}, \bet^{\LR}_{k+1} \mid \bz^{\LR}_{1:k})$ is estimated via two posterior densities as (see~\cite{Alotaibi-2024_TL_CKF_arXiv})
\be\label{eq:BTL_sr_overall}
p(\bx^{\LR}_{k}, \bet^{\LR}_{k+1} \mid \bz^{\LR}_{1:k}) \propto\  p(\bx^{\LR}_{k} \mid \bz^{\LR}_{1:k})\, p(\bet^{\LR}_{k+1} \mid \bz^{\LR}_{1:k}) ,
\ee
where the object state posterior and predicted observation densities are expressed, respectively, as 
\begin{align}\label{eq:BTL_sr_state_1}
p(\bx^{\LR}_{k} \mid \bz^{\LR}_{1:k}) 
\propto\ p(\bz^{\LR}_{k} \mid \bx^{\LR}_{k})\, p(\bx^{\LR}_{k} \mid \bx^{\LR}_{k-1})\, p(\bx^{\LR}_{k-1} \mid \bz^{\LR}_{1:k-1}) ,
\end{align}
and
\begin{align}\label{eq:BTL_sr_eta_1}
p(\bet^{\LR}_{k+1} \mid \bz^{\LR}_{1:k}) \propto\ 
p(\bet^{\LR}_{k+1} \mid \bx^{\LR}_{k+1})\, p(\bx^{\LR}_{k+1} \mid \bx^{\LR}_{k})\,
p(\bx^{\LR}_{k} \mid \bz^{\LR}_{1:k}) .
\end{align}
The parameter of the predicted observation density in~\eqref{eq:BTL_sr_eta_1}
along with the predicted observations are transferred to the primary tracking filter and leveraged in the tracking filter of the next time step to improve tracking accuracy.

\subsection{Primary Tracking Filter}
The primary tracking filter receives predicted observations up to time step $k$ from the source tracking filter. Given the set of transferred predicted observations $\bet^{\LR}_{2:k}$ and observed measurements up to time step $k$, the object state posterior density is estimated as (see~\cite{Alotaibi-2024_TL_CKF_arXiv})
\begin{align} \label{eq:BTL_tr_overall_2}
	p(\bx^{\TR}_{k} \mid \bz^{\TR}_{1:k}, \bet^{\LR}_{2:k})
 \propto\ 
p( \bz^{\TR}_{k} \mid \bx^{\TR}_{k})\  p(\bet^{\LR}_{k} \mid \bx^{\TR}_{k}) \ p(\bx^{\TR}_{k} \mid \bx^{\TR}_{k-1}) .
\end{align}
Via~\eqref{eq:BTL_tr_overall_2}, the state density is estimated by incorporating transferred predicted observations from the source tracking filter as prior knowledge. This yields superior performance for the primary sensor, which has
less reliable measurements.

%% file: PF_TL_Section.tex
\section{Transfer Learning for Particle Filters} \label{Sec:TL_PF}

We now develop a global approximation approach that integrates
particle filtering with a \ac{btlf} to approximate posterior densities of the source and primary tracking filters and to leverage the knowledge transferred between tracking filters. In this section, \ac{sir-pf} is considered to be the global approach to approximate, which offers improved estimation accuracy compared to local approaches by relaxing assumptions on the forms of prior and posterior densities.

%the global approximation approach, \ac{sir-pf}, can provide a better approximation with less strict assumption of prior and posterior forms and able to achieve higher levels of accuracy estimation compared to local approaches. 

\subsection{Source Tracking Filter}

The object state posterior density $p(\bx^{\LR}_{k} \mid \bz^{\LR}_{1:k})$ in~\eqref{eq:BTL_sr_state_1} is approximated using \ac{pf} with $N_s$ weighted particles given by
\be\label{eq:BTL_sr_state_2}
p(\bx^{\LR}_{k} \mid \bz^{\LR}_{1:k}) \approx \sum_{i=1}^{N_s} \ w_{k}^{\LR (i)} \delta \left( \bx^{\LR}_{k} - \bx^{\LR (i)}_{k}\right) ,
\ee
where $\delta(\cdot)$ is the Dirac delta function. Ideally, the particles in~\eqref{eq:BTL_sr_state_2} are sampled directly from the posterior density. Since sampling from the posterior density itself is not possible in most practical scenarios, weighted particles are simply drawn indirectly from a proposal density, which is known as the importance density and derived in Appendix~\ref{TL_PF_Derivation_Section}, defined as 
\be\label{eq:BTL_sr_particle_1}
\bx^{\LR (i)}_{k} \sim q(\bx^{\LR (i)}_{k} \mid \bx^{\LR (i)}_{k-1}, \bz^{\LR}_{1:k})\bigg\rvert_{i=1,...,N_s} .
\ee

The importance density $q(\bx^{\LR (i)}_{k} \mid \bx^{\LR (i)}_{k-1}, \bz^{\LR}_{1:k}) $ is chosen to be the state transition prior $p(\bx^{\LR (i)}_{k} \mid \bx^{\LR (i)}_{k-1})$ given in~\eqref{Pro_form_State}. This choice of proposal density was used in~\cite{Liu-1998-PF_N_effective,ArulaMGC-2002-Tutorial_PF_EKF} to simplify the implementation of 
the \ac{pf}. According to this common choice of proposal density, the computation time for sampling from the proposal importance density will be eliminated by propagating samples of the previous time step through the transition model function. After choosing the importance density as the state transition prior, the non-normalized weights in \eqref{eq:BTL_sr_state_2} and sampled particles in~\eqref{eq:BTL_sr_particle_1} can be obtained, respectively, as
\be\label{eq:TL_sr_weights_1}
w_{k}^{\LR (i)} \propto w_{k-1}^{\LR (i)}\ p(\bz^{\LR}_{k} \mid \bx^{\LR (i)}_{k}),
\ee
and
\be\label{eq:BTL_sr_particle_2}
\bx^{\LR (i)}_{k} \sim p(\bx^{\LR (i)}_{k} \mid \bx^{\LR (i)}_{k-1})\bigg\rvert_{i=1,...,N_s} .
\ee

Under the assumption $\bw^{\LR}_{k} \stackrel{\rm iid}{\sim} \ccN (\mathbf{0}, \bQ^{\LR}_{\bw})$, the measurement likelihood is utilized to update non-normalized weights in~\eqref{eq:TL_sr_weights_1} via 
\be
p(\bz^{\LR}_{k} \mid \bx^{\LR (i)}_{k}) = \frac{1}{(2\pi)^{n_\bz/2}\ \sqrt{|\bQ^{\LR}_{\bw}|}}\ \exp \left( \rho^{\LR (i)} \right) ,
\label{eq:TL_sr_likelihood_1}
\ee
where 
\be
\rho^{\LR (i)} = -\frac{1}{2} \left(\bz^{\LR}_{k} \! - \! h(\bx^{\LR (i)}_{k}) \right)^{T}  (\bQ^{\LR}_{\bw})^{-1}  \left(\bz^{\LR}_{k} \! -  \! h(\bx^{\LR (i)}_{k}) \right) ,
\label{eq:TL_sr_likelihood_2}
\ee
$\bz^{\LR}_k \in \mathbb{R}^{n_\bz}$, and $\bQ^{\LR}_{\bw}$ is the covariance of the Gaussian measurement noise in~\eqref{Pro_form_meas}. The updated weights in~\eqref{eq:TL_sr_weights_1} are normalized via
\be
\tilde{w}_{k}^{\LR (i)} = \frac{w_{k}^{\LR (i)}} {\sum_{i=1}^{N_s} w_{k}^{\LR (i)}} .
\ee

The resampling step reduces the effect of the degeneracy phenomenon described in~\cite{DouceGA-2000-PF_Degeneracy}, which appears after performing the sequential algorithm for a certain number of iterations. This phenomenon causes most particles to receive near-zero weight, rendering them ineffective in the approximation process for the posterior density. This method measures the degeneracy at each time step by computing the effective number of particles $N_{eff}$, as defined in~\cite{Liu-1998-PF_N_effective}, and plays an important role once the effective number of particles drops below a certain threshold. The systematic resampling technique introduced in~\cite{Kitagawa-1996-PF_Systematic_Resampling} is performed at every time step in the \ac{sir} by neglecting particles with low weights and duplicating the
higher weighted particles. The new set of particles is re-weighted uniformly by assigning a new weight $w_{k}^{\LR (i)} = 1/N_{s}$ to each particle as a part of the \ac{sir} process.

The state posterior density, $p(\bx^{\LR}_{k} \mid \bz^{\LR}_{1:k}) = \ccN (\bx^{\LR}_{k}; \bhx^{\LR}_{k} , \bP^{\LR}_{k})$,  is approximately estimated based on the re-sampled particles and new assigned weights with mean $\bhx^{\LR}_{k}$ computed as
\be\label{eq:BTL_sr_state_mean}
\bhx^{\LR}_{k} = \frac{1}{N_{s}} \sum_{i=1}^{N_{s}}\ \bx^{\LR (i)}_{k} ,
\ee
and the associated covariance $\bP^{\LR}_{k}$ is obtained via
\begin{align}
\bP_{k}^{\LR} = \frac{1}{N_{s}} \sum_{i=1}^{N_{s}}\
\left( \bx^{\LR (i)}_{k}  - \bhx^{\LR}_{k} \right) \left( \bx^{\LR (i)}_{k} - \bhx^{\LR}_{k} \right)^{T} + \bQ_{\bv}^{\LR} . 
\label{eq:BTL_sr_state_cov}
\end{align}

The predicted observation posterior density $p(\bet^{\LR}_{k+1} \mid \bz^{\LR}_{1:k})$ in~\eqref{eq:BTL_sr_eta_1} is approximated similarly with a set of weighted particles via 
\be\label{eq:BTL_sr_pred_obser_1}
p(\bet^{\LR}_{k+1} \mid \bz^{\LR}_{1:k}) \approx \sum_{i=1}^{N_s} \ w_{k+1}^{\LR \bet (i)} \delta \left( \bet^{\LR}_{k+1} - \bet^{\LR (i)}_{k+1}\right) ,
\ee
where the importance density $q(\bet^{\LR  (i)}_{k+1} \mid \bx^{\LR (i)}_{k}, \bz^{\LR}_{1:k})$, a proof of derivation is provided in Appendix~\ref{TL_PF_Derivation_Section}, is chosen to be $p(\bet^{\LR (i)}_{k+1} \mid \bx^{\LR (i)}_{k})$ given as
\be\label{eq:BTL_sr_importance_1}
p(\bet^{\LR (i)}_{k+1} \mid \bx^{\LR (i)}_{k}) = p(\bet^{\LR (i)}_{k+1} \mid \bx^{\LR (i)}_{k+1})\ p(\bx^{\LR (i)}_{k+1} \mid \bx^{\LR (i)}_{k}) .
\ee
 According to the choice of importance density in~\eqref{eq:BTL_sr_importance_1}, particles are drawn and weighted as following:
\begin{align}
\label{eq:BTL_sr_particle_3}
\bx^{\LR (i)}_{k+1} & \sim p(\bx^{\LR (i)}_{k+1} \mid \bx^{\LR (i)}_{k})\bigg\rvert_{i=1,...,N_s} , \\
\label{eq:BTL_sr_particle_4}
\bet^{\LR (i)}_{k+1} &\sim p(\bet^{\LR (i)}_{k+1} \mid \bx^{\LR (i)}_{k+1})\bigg\rvert_{i=1,...,N_s} ,
\end{align}
and
\be\label{eq:TL_sr_weights_2}
w_{k+1}^{\LR \bet (i)} =  w_{k}^{\LR (i)} = 1/N_{s} .
\ee
%$w_{k+1}^{\LR \bet (i)} =  w_{k}^{\LR (i)} = 1/N_{s}$. 

The mean and covariance for the predicted observation posterior density in~\eqref{eq:BTL_sr_pred_obser_1}, which follows a Gaussian model expressed as $p(\bet^{\LR}_{k+1} \mid \bz^{\LR}_{1:k})= \ccN (\bet^{\LR}_{k+1}; \bhet^{\LR}_{k+1}, \bP^{\LR}_{\bet\bet,k+1})$, are estimated via uniformly weighted sampled particles and computed according to the following expressions:
\begin{align}
\bhet^{\LR}_{k \! + \! 1} & = \frac{1}{N_{s}} \sum_{i=1}^{N_{s}}\ \bet^{\LR (i)}_{k+1} , \label{eq:BTL_sr_pred_obser_mean} \\
\bP^{\LR}_{\bet\bet,k \! + \! 1} 
 \! & =
 \frac{1}{N_{s}} \sum_{i=1}^{N_{s}} \!
\left( \bet^{\LR (i)}_{k+1} \! - \! \bhet^{\LR}_{k+1} \right) \left( \bet^{\LR (i)}_{k+1} \! - \! \bhet^{\LR}_{k+1} \right)^{T} \! + \! \bQ_{\bw}^{\LR} . 
\label{eq:BTL_sr_pred_obser_cov}
\end{align}
Note that the estimated mean $\bhet^{\LR}_{k+1} $ and covariance $\bP^{\LR}_{\bet\bet,k+1} $ in~\eqref{eq:BTL_sr_pred_obser_mean} and~\eqref{eq:BTL_sr_pred_obser_cov}, respectively, are transferred simultaneously to the primary tracking filter. These transferred parameters,
which characterize the predicted observation density in~\eqref{eq:BTL_sr_eta_1}, provide valuable knowledge to be leveraged in the estimation process of the primary tracking filter
to enhance its tracking performance. The proposed source tracking filter approach is summarized with procedural steps in Algorithm~\ref{alg_TL_PF_LR}.

\input{alg_TL_PF_LR}

\subsection{Primary Tracking Filter}

The primary sensor utilizes the transferred mean and covariance of the predicted observation density from the source tracking filter as a prior 
to estimate the overall posterior density $p(\bx^{\TR}_{k} \mid \bz^{\TR}_{1:k}, \bet^{\LR}_{2:k})$ given in~\eqref{eq:BTL_tr_overall_2}. Using the~\ac{pf} approach, the approximation of the overall posterior density via weighted particles is expressed as 
\be\label{eq:BTL_tr_state_2}
p(\bx^{\TR}_{k} \mid \bz^{\TR}_{1:k}, \bet^{\LR}_{2:k}) \approx \sum_{i=1}^{N_s} \ w_{k}^{\TR (i)} \delta \left( \bx^{\TR}_{k} - \bx^{\TR (i)}_{k}\right) ,
\ee
where the particles in~\eqref{eq:BTL_tr_state_2} are drawn from an importance density, which is derived in Appendix~\ref{TL_PF_Derivation_Section}, as follows:
\be\label{eq:BTL_tr_particle_1}
\bx^{\TR (i)}_{k} \sim q(\bx^{\TR (i)}_{k} \mid \bx^{\TR (i)}_{k-1}, \bz^{\TR}_{1:k}, \bet^{\LR}_{2:k})\bigg\rvert_{i=1,...,N_s} .
\ee

By choosing the proposal density to be the state transition prior, i.e.,
$q(\bx^{\TR (i)}_{k} \mid \bx^{\TR (i)}_{k-1}, \bz^{\TR}_{1:k}, \bet^{\LR}_{2:k}) = p(\bx^{\TR (i)}_{k} \mid \bx^{\TR (i)}_{k-1})$, the non-normalized weights in~\eqref{eq:BTL_tr_state_2} are computed via
\be\label{eq:TL_tr_weights_1}
w_{k}^{\TR (i)} \propto w_{k}^{\TR \bet (i)} \ p(\bz^{\TR}_{k} \mid \bx^{\TR (i)}_{k}),
\ee
where the transferred predicted observation weights are denoted by $w_{k}^{\TR \bet (i)}$ and defined as 
\be\label{eq:TL_tr_weights_2}
w_{k}^{\TR \bet (i)} \propto w_{k-1}^{\TR (i)} \ p(\bet^{\LR}_{k} \mid \bx^{\TR (i)}_{k}),
\ee
and particles that approximate the overall posterior density, formulated in~\eqref{eq:BTL_tr_particle_1}, are sampled based on the chosen proposal density as 
\be\label{eq:BTL_tr_particle_2}
\bx^{\TR (i)}_{k} \sim p(\bx^{\TR (i)}_{k} \mid \bx^{\TR (i)}_{k-1})\bigg\rvert_{i=1,...,N_s} .
\ee

The non-normalized weights are updated using two likelihoods to incorporate the most recent observed measurement and the simultaneously transferred parameters at each time step. The transferred predicted observation weights in~\eqref{eq:TL_tr_weights_2} are obtained as an initial stage for updating the overall weights in~\eqref{eq:TL_tr_weights_1} by applying the transferred predicted observation likelihood $p(\bet^{\LR}_{k} \mid \bx^{\TR (i)}_{k})$,
which is given by
\be
p(\bet^{\LR}_{k} \mid \bx^{\TR (i)}_{k}) = \frac{1}{(2\pi)^{n_\bz/2}\ \sqrt{|\bP^{\LR}_{\bet\bet,k}|}}\ \exp \left( \rho^{\TR \bet (i)} \right) ,
\label{eq:TL_tr_likelihood_1}
\ee
where
\begin{align}
 \rho^{\TR \bet (i)} = -\frac{1}{2} \left(\bhet^{\LR}_{k} - h(\bx^{\TR (i)}_{k}) \right)^{T}  (\bP^{\LR}_{\bet\bet,k})^{-1}  \left(\bhet^{\LR}_{k} - h(\bx^{\TR (i)}_{k}) \right) .
\label{eq:TL_tr_likelihood_2}
\end{align}

After updating the weights through the transferred predicted observation likelihood, the measurement likelihood $p(\bz^{\TR}_{k} \mid \bx^{\TR (i)}_{k})$ is employed to compute the weights in~\eqref{eq:TL_tr_weights_1}. The measurement likelihood is given by
\be
p(\bz^{\TR}_{k} \mid \bx^{\TR (i)}_{k}) = \frac{1}{(2\pi)^{n_\bz/2}\ \sqrt{|\bQ^{\TR}_{\bw}|}}\ \exp \left( \rho^{\TR (i)} \right) ,
\label{eq:TL_tr_likelihood_3}
\ee
where
\be
\rho^{\TR (i)} = -\frac{1}{2} \left(\bz^{\TR}_{k} - h(\bx^{\TR (i)}_{k}) \right)^{T} ( \bQ^{\TR}_{\bw})^{-1}  \left(\bz^{\TR}_{k} - h(\bx^{\TR (i)}_{k}) \right) .
\label{eq:TL_tr_likelihood_4}
\ee

The above likelihoods, expressed in~\eqref{eq:TL_tr_likelihood_1} and~\eqref{eq:TL_tr_likelihood_3}, are formulated under the assumption of the measurement model given by~\eqref{Pro_form_meas}. 
Unlike the isolated~\ac{pf} algorithm, \ac{tl-pf} updates particle weights using transferred density parameters. These include the mean $\bhet^{\LR}_{k}$ and its associated covariance $\bP^{\LR}_{\bet\bet,k}$, which are estimated at the previous time step $k-1$ from the source tracking filter as an additive step introduced mainly for incorporating transferred parameters within the process of the~\ac{pf} framework in addition to the updating step in~\eqref{eq:TL_tr_likelihood_3} and~\eqref{eq:TL_tr_likelihood_4} via newly observed measurement $\bz^{\TR}_{k}$ with its covariance $\bQ^{\TR}_{\bw}$. The overall weights associated with each individual particle in~\eqref{eq:TL_tr_weights_1} are normalized by
\be
\tilde{w}_{k}^{\TR (i)} = \frac{w_{k}^{\TR (i)}} {\sum_{i=1}^{N_s} w_{k}^{\TR (i)}} .
\ee

Similar to the source tracking filter, drawn particles along with their normalized weights are resampled using the systematic resampling technique to mitigate the degeneracy phenomenon effect where the resultant weights are identically equal to $1/N_{s}$. The overall state posterior density in~\eqref{eq:BTL_tr_state_2} is modeled as a Gaussian density, expressed as $p(\bx^{\TR}_{k} \mid \bz^{\TR}_{1:k}, \bet^{\LR}_{2:k}) = \ccN (\bx^{\TR}_{k}; \bhx^{\TR}_{k} , \bP^{\TR}_{k})$. The mean $\bhx^{\TR}_{k}$ and associated covariance $\bP^{\TR}_{k}$ of this posterior density are  estimated using $N_{s}$ resampled particles as follows:
\begin{align} \label{eq:BTL_tr_state_mean}
\bhx^{\TR}_{k} &= \frac{1}{N_{s}} \sum_{i=1}^{N_{s}}\ \bx^{\TR (i)}_{k} , \\
\bP_{k}^{\TR} &= \frac{1}{N_{s}} \sum_{i=1}^{N_{s}}\
\left( \bx^{\TR (i)}_{k}  - \bhx^{\TR}_{k} \right) \left( \bx^{\TR (i)}_{k} - \bhx^{\TR}_{k} \right)^{T} + \bQ_{\bv}^{\TR} . 
\label{eq:BTL_tr_state_cov}
\end{align}

Algorithm~\ref{alg_TL_PF_TR} provides detailed steps of the proposed primary tracking filter. The \ac{tl-pf} algorithm describes an implementation for approximating the \ac{btlf} using the \ac{pf} specifically \ac{sir} method by leveraging transferred knowledge from the source tracking filter into the primary tracking filter.

\input{alg_TL_PF_TR}

%% file: alg_TL_PF_LR.tex
\begin{algorithm}%[H]
\caption{TL-PF Source Tracking Algorithm}
\label{alg_TL_PF_LR}
{%
\linespread{1.2}\selectfont

\textbf{Input:} $\bz_{k}^{\LR}$, $\bx_{k}^{\LR (i)}$, $i = 1, \ldots, N_s$ \\
\textbf{Output:} $\bhx^{\LR}_{k}$, $\bhet^{\LR}_{k+1}$, $\bP^{\LR}_{\bet\bet,k+1}$, $\bx_{k+1}^{\LR (i)}$, $i = 1, \ldots, N_s$ \\

\begin{algorithmic}[1]

\vsp{.1in}

\For{particles, $i \gets 1$ to $N_s$}
\vsp{.02in}
    \State{Obtain measurement likelihood  $p(\bz^{\LR}_{k} \mid \bx^{\LR (i)}_{k})$\Comment{\eqref{eq:TL_sr_likelihood_1}}} 
    \State{Compute weight $w_{k}^{\LR (i)}$\Comment{\eqref{eq:TL_sr_weights_1}\vsp{.02in}}}
%
%\vsp{.02in}
%
\EndFor
\vsp{.1in}
\State{Normalize weights $\tilde{w}_{k}^{\LR (i)}  = \frac{w_{k}^{\LR (i)}} {\sum_{i=1}^{N_s} w_{k}^{\LR (i)}}$\vsp{.05in}} 
\vsp{.02in}
\State{$\left\lbrace \bx_{k}^{\LR (i)}, w_{k}^{\LR (i)} \right\rbrace\!=\!$ Resample $\left\lbrace \bx_{k}^{\LR (i)}, \tilde{w}_{k}^{\LR (i)} \right\rbrace$, $i = 1, \ldots, N_s$\vsp{.1in}}
%
%\vsp{.05in}
%
\State{Compute state posterior $p(\bx^{\LR}_{k} \mid \bz^{\LR}_{1:k})$ \Comment{\eqref{eq:BTL_sr_state_mean}, \eqref{eq:BTL_sr_state_cov}}\vsp{.1in}}
%
%
%\vsp{.01in}
%
%
\For{particles, $i \gets 1$ to $N_s$}
    \State{Draw predicted state particle $\bx_{k+1}^{\LR (i)}$ \Comment{\eqref{eq:BTL_sr_particle_3}}}
    \State{Draw predicted observation particle $\bet^{\LR (i)}_{k+1}$ \Comment{\eqref{eq:BTL_sr_particle_4}}}
    \State{Compute associated weight $w_{k+1}^{\LR \bet (i)}$ \Comment{\eqref{eq:TL_sr_weights_2}}}
\EndFor
\vsp{.1in}
\State{Compute predicted observation posterior

$p(\bet^{\LR}_{k+1} \mid \bz^{\LR}_{1:k})$~\Comment{\eqref{eq:BTL_sr_pred_obser_mean}, \eqref{eq:BTL_sr_pred_obser_cov}\vsp{.1in}}}
%
%\vsp{.1in}
%

\vsp{.1in}

\end{algorithmic}

\textbf{Transfer:} the estimated density parameters $\bhet^{\LR}_{k+1}$ and $\bP^{\LR}_{\bet\bet,k+1}$ to the primary tracking filter.

    }
\end{algorithm}

%% file: alg_TL_PF_TR.tex
\begin{algorithm}%[H]
\caption{TL-PF Primary Tracking Algorithm}
\label{alg_TL_PF_TR}
{%
\linespread{1.2}\selectfont

\textbf{Input:} $\bz_{k}^{\TR}$, $\bhet^{\LR}_{k}$, $\bP^{\LR}_{\bet\bet,k}$, $\bx_{k-1}^{\TR (i)}$, $i = 1, \ldots, N_s$ \\
\textbf{Output:} $\bhx^{\TR}_{k}$, $\bx_{k}^{\TR (i)}$, $i = 1, \ldots, N_s$ \\

\begin{algorithmic}[1]

\vsp{.1in}

\For{particles, $i \gets 1$ to $N_s$}
    \State{Draw state particle $\bx^{\TR (i)}_{k}$ \Comment{\eqref{eq:BTL_tr_particle_2}}}
%
%\State{\txxx  Obtain transferred predicted observation likelihood $p(\bet^{\LR}_{k} \mid \bx^{\TR (i)}_{k})$ via \eqref{eq:TL_tr_likelihood_1}.}
%
    \State{Calculate TL likelihood $p(\bet^{\LR}_{k} \mid \bx^{\TR (i)}_{k})$ \Comment{\eqref{eq:TL_tr_likelihood_1}}}
    \State{Obtain TL weight $w_{k}^{\TR \bet (i)}$ \Comment{\eqref{eq:TL_tr_weights_2}}}
    \State{Compute measurement likelihood  $p(\bz^{\TR}_{k} \mid \bx^{\TR (i)}_{k})$ \Comment{\eqref{eq:TL_tr_likelihood_3}}}
    \State{Obtain weight $w^{\TR (i)}_{k}$ \Comment{\eqref{eq:TL_tr_weights_1}}}
\EndFor
\vsp{.1in}
\State{Normalize weights $\tilde{w}_{k}^{\TR (i)} = \frac{w_{k}^{\TR (i)}} {\sum_{i=1}^{N_s} w_{k}^{\TR (i)}}$\vsp{.05in}} 
\State{$\left\lbrace \bx_{k}^{\LR (i)}, w_{k}^{\LR (i)} \right\rbrace\!=\!$ Resample $\left\lbrace \bx_{k}^{\LR (i)}, \tilde{w}_{k}^{\LR (i)} \right\rbrace$, $i = 1, \ldots, N_s$\vsp{.05in}}
\State{Estimate overall state posterior 

$p(\bx^{\TR}_{k} \mid \bz^{\TR}_{1:k}, \bet^{\LR}_{2:k})$ \Comment{\eqref{eq:BTL_tr_state_mean}, \eqref{eq:BTL_tr_state_cov}}}

	\end{algorithmic}
    }
\end{algorithm}

%% file: PF_Sim_v2.tex
\section{Simulation Results}
\label{Sec:TL_PF_Sim}

\subsection{Scenario 1: Linear Motion Model}
\subsubsection{Tracking System Model and Parameters Settings}

In Scenario~1, we consider a single object following a linear motion model described as
\begin{align} \label{eq:linear_model_sim}
\bx_{k} &= 
\begin{bmatrix}
	1 & T_{s} & 0 & 0 \\
	0 & 1 & 0 & 0 \\
	0 & 0 & 1 & T_{s}\\
	0 & 0 & 0 & 1 
\end{bmatrix}
\bx_{k-1} + \bv_{k-1},
\end{align}
where the unknown object state is defined as $\bx_{k} \eqq [ x_{k}, \dot{x}_{k}, y_{k}, \dot{y}_{k}]^T$ consisting of the object's position and velocity denoted by $(x_{k}, y_k)$ and $(\dot{x}_{k}, \dot{y}_{k})$, respectively. The process noise is a zero-mean Gaussian noise with covariance given by
\be
\bQ_{\bv}=
\begin{bmatrix}
	q\frac{T_{s}^{4}}{4} & q\frac{T_{s}^{3}}{2} & 0 & 0 \\
	q\frac{T_{s}^{3}}{2} & q T_{s}^{2}& 0 & 0 \\
	0 & 0 & q \frac{T_{s}^{4}}{4}& q\frac{T_{s}^{3}}{2} \\
	0 & 0 & q \frac{T_{s}^{3}}{2}  & q T_{s}^{2}
\end{bmatrix} .
\ee

\begin{figure}
	\centering
	\includegraphics[trim={1.85cm 6.8cm 1.65cm 7.5cm},clip, scale=0.7]
        {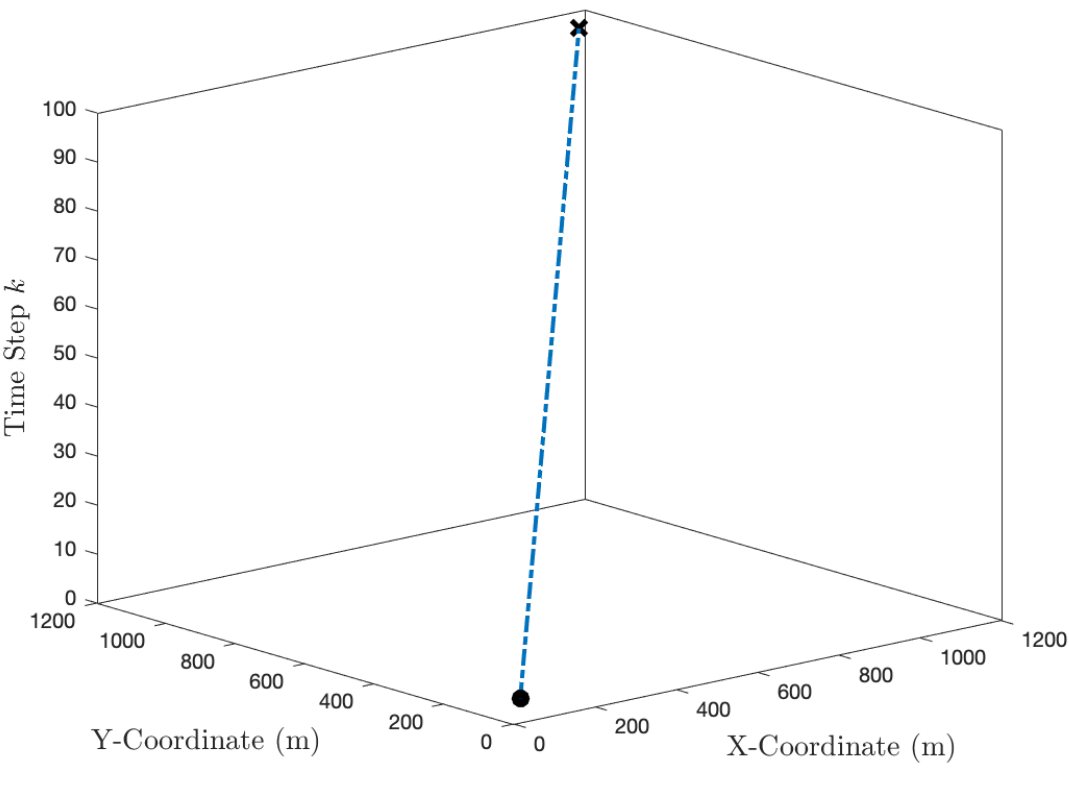}
	\caption{Scenario 1: A moving object trajectory in two-dimensional coordinates with an initial point indicated by $\sbullet[1.2]$.}
	\label{fig:Linear_tracjectory_3D}
\end{figure}

The initialization of the object's state and covariance in Scenario~1 are 
\begin{align}
\bx_{0}&=[100~\mathrm{m}, 10~\mathrm{m/s}, 100~\mathrm{m}, 10~\mathrm{m/s}]^{T} , \\
\bP_{0}&=\text{diag}[50~\mathrm{m^2}, 1~\mathrm{m^2/s^2}, 50~\mathrm{m^2}, 1~\mathrm{m^2/s^2}] . 
\end{align} 
The source and primary tracking filters observe the measurements that follow the measurement models in \eqref{Pro_form_meas_LR} and \eqref{Pro_form_meas_TR} with asymmetric measurement noise intensities. The measurement vector for each sensor comprises the object's range, $r_k \eqq \sqrt{x_{k}^{2} + y_{k}^{2}}$, and bearing angle, $\zeta_k \eqq \arctan\left(y_{k}/x_{k}\right)$, i.e., $\bz_k = [r_k, \zeta_k]^T$. The measurement noises of the source and primary tracking filters are zero-mean Gaussian denoted as $\bw^{\LR}_{k} \stackrel{\rm iid}{\sim} \ccN (\mathbf{0}, \bQ^{\LR}_{\bw})$ and $\bw_{k} \stackrel{\rm iid}{\sim} \ccN (\mathbf{0}, \bQ_{\bw})$, with associated covariances $\bQ^{\LR}_{\bw}=I^{\LR}_{\bw}\ \bB_\bw$ and $\bQ_{\bw}=I_{\bw}\ \bB_\bw$, respectively,  where $\bB_\bw=\text{diag}[\sigma_{r}^{2},  \sigma_{\zeta}^{2}]$. Noise intensity levels $I^{\LR}_{\bw}$ and $I^{\TR}_{\bw}$ represent conditions impacting the individual sensors. The simulation parameters in Scenario~1 are set as $n_\bx = 4$, $n_\bz = 2$, $T_s = 1\ \mathrm{s}$, $q = 0.1\ \mathrm{m^2/s^4}$, $\sigma_{r} = 10\ \mathrm{m}$, $\sigma_{\zeta} = \sqrt{10}\times 10^{-3}\ \mathrm{rad}$, $I^{\LR}_{\bw} = 1$, and $I^{\TR}_{\bw} = 4$. The trajectory of the moving object that follows the linear motion model in~\eqref{eq:linear_model_sim} for $100$ time steps, as shown in Fig.\ref{fig:Linear_tracjectory_3D}.

\subsubsection{Performance Evaluation}

\begin{figure}
	\centering
	\includegraphics[trim={2cm 6.8cm 2cm 7cm},clip, scale=0.7]{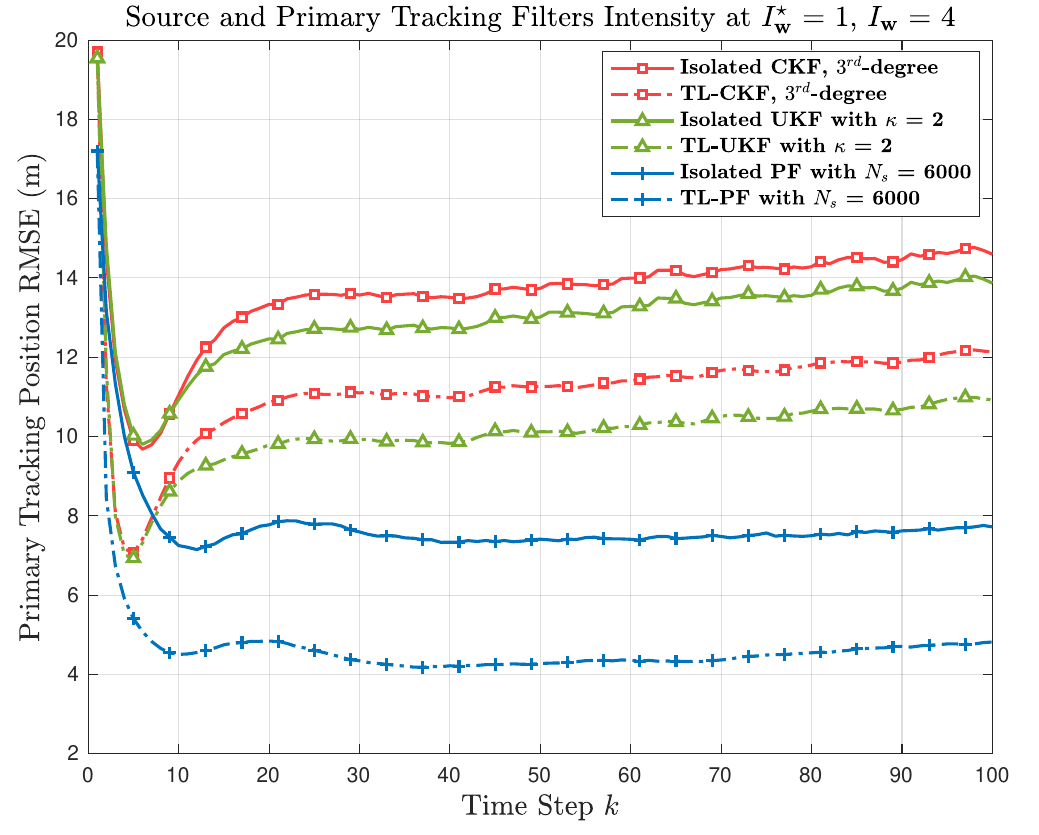}
	\caption{Scenario 1: \ac{rmse} values plots computed via~\eqref{eq:RMSE_per_k} for the proposed \ac{tl-pf} (blue line with $\sadd$ marks) algorithm along with isolated and \ac{btlf} of \ac{ukf} (green line with $\striu$~marks) and third-degree \ac{ckf} (red line with $\ssqur$~marks).}
	\label{fig:Linear_PF_RMSE_vs_Time_6000}
\end{figure}

We evaluate the proposed method for Scenario~1 that provided in Fig.\ref{fig:Linear_tracjectory_3D} via the \ac{rmse} of the object's position for the primary tracking filter. The \ac{rmse} metric is computed by averaging $10,000$ \ac{mc} simulation iterations at time step $k$ as:
\begin{align} \label{eq:RMSE_per_k}
\text{RMSE} (k) = \sqrt{\frac{1}{MC} \sum_{m=1}^{MC} (\text{True\ Pos.}_{m,k} - \text{Est.\ Pos.}_{m,k})^2} .
\end{align}
Figure~\ref{fig:Linear_PF_RMSE_vs_Time_6000} shows \ac{rmse} results for \ac{tl-pf} ($N_s = 6000$ particles) compared with \ac{tl-ukf} and third-degree \ac{tl-ckf}. As shown in Fig.~\ref{fig:Linear_PF_RMSE_vs_Time_6000}, the proposed \ac{tl-pf} algorithm (blue dashed line with $\sadd$ marks) has lower \ac{rmse} values at the entire duration of the tracked trajectory compared to the isolated \ac{pf} (blue solid line with $\sadd$ marks). For instance, the \ac{tl-pf} is able to achieve \ac{rmse} value of $4.28$~m at time step $k=52$~s, while the isolated \ac{pf} has \ac{rmse} value of $7.37$~m at the same time step. Compared with previously proposed algorithms, the \ac{tl-ukf} (green dashed line with $\striu$~marks) and third-degree \ac{tl-ckf} (red dashed line with $\ssqur$~marks) achieve \ac{rmse} values of $10.13$~m and $11.28$~m, respectively, at the same time step $k=52$~s. For the entire duration of the object trajectory in Scenario~1, the \ac{tl-pf} algorithm performs better with lower \ac{rmse} values compared to comparable algorithms with \ac{btlf} and isolated (without incorporation of the transfer learning approach).

\subsection{Scenario 2: Nonlinear Maneuvering Motion Model}
\subsubsection{Tracking System Model}

For our simulations, we consider a single maneuvering object with an unknown state vector  to be tracked by a dual-tracking system. We adopt the nonlinear motion model in~\cite{Shalom-2001-EAT} given  by
\begin{align}
\bx_{k} &= 
\begin{bmatrix}
	1 & \frac{\sin (\Omega_{k-1} T_{s})}{\Omega_{k-1}} & 0 & -\left(\frac{1-\cos (\Omega_{k-1} T_{s})}{\Omega_{k-1}}\right) & 0\\
	0 & \cos (\Omega_{k-1} T_{s}) & 0 & -\sin (\Omega_{k-1} T_{s}) & 0\\
	0 & \frac{1-\cos (\Omega_{k-1} T_{s})}{\Omega_{k-1}} & 1 &  \frac{\sin (\Omega_{k-1} T_{s})}{\Omega_{k-1}} & 0\\
	0 & \sin (\Omega_{k-1} T_{s}) & 0 & \cos (\Omega_{k-1} T_{s})  & 0\\
	0 & 0 & 0 & 0 & 1
\end{bmatrix}
\bx_{k-1} \nonumber \\
&~~~~~~+ \bv_{k-1} .
\end{align}
The object state is given by $\bx_{k} \eqq [ x_{k}, \dot{x}_{k}, y_{k}, \dot{y}_{k}, \Omega_{k}]^T$,
which consists of the Cartesian coordinates of the object's position $(x_{k}, y_{k})$, the object's velocity $(\dot{x}_{k}, \dot{y}_{k})$, and the turn rate $\Omega_{k}$. The process noise $\bv_{k-1} \stackrel{\rm iid}{\sim} \mathcal{N}(0,\bQ_{\bv})$ with associated covariance 
\be
\bQ_{\bv}=
\begin{bmatrix}
	q_{1}\frac{T_{s}^{4}}{4} & q_{1}\frac{T_{s}^{3}}{2} & 0 & 0 & 0\\
	q_{1}\frac{T_{s}^{3}}{2} & q_{1}T_{s}^{2}& 0 & 0 & 0 \\
	0 & 0 & q_{1}\frac{T_{s}^{4}}{4}& q_{1}\frac{T_{s}^{3}}{2} & 0 \\
	0 & 0 & q_{1}\frac{T_{s}^{3}}{2}  & q_{1}T_{s}^{2} & 0\\
	0 & 0 & 0 & 0 & q_{2}T_{s}
\end{bmatrix} .
\ee
As in Section~\ref{Sec:Problem_form}, the error process noise covariances of the source and primary tracking filters are assumed to be identical.

\subsubsection{Parameters Settings}

%The dual-sensor system observes measurements that follow the nonlinear measurement models in~\eqref{Pro_form_meas_TR} with asymmetric measurement noise intensities for the source and primary tracking filters. The measurement vector for each sensor comprises the object's range, $r_k \eqq \sqrt{x_{k}^{2} + y_{k}^{2}}$, and bearing angle, $\zeta_k \eqq \arctan\left(y_{k}/x_{k}\right)$, i.e., $\bz_k = [r_k, \zeta_k]^T$. The measurement noises of the source and primary tracking filters are zero-mean Gaussian denoted as  $\bw^{\LR}_{k} \stackrel{\rm iid}{\sim} \ccN (\mathbf{0}, \bQ^{\LR}_{\bw})$ and $\bw_{k} \stackrel{\rm iid}{\sim} \ccN (\mathbf{0}, \bQ_{\bw})$, with associated covariances $\bQ^{\LR}_{\bw}=I^{\LR}_{\bw}\ \bB_\bw$ and $\bQ_{\bw}=I_{\bw}\ \bB_\bw$, respectively,  where $\bB_\bw=\text{diag}[\sigma_{r}^{2},  \sigma_{\zeta}^{2}]$. The noise intensity levels $I^{\LR}_{\bw}$ and $I^{\TR}_{\bw}$ represent conditions impacting the individual sensors. 

\begin{figure}
	\centering
	\includegraphics[trim={1.85cm 6.8cm 1.65cm 7.5cm},clip, scale=0.7]
        {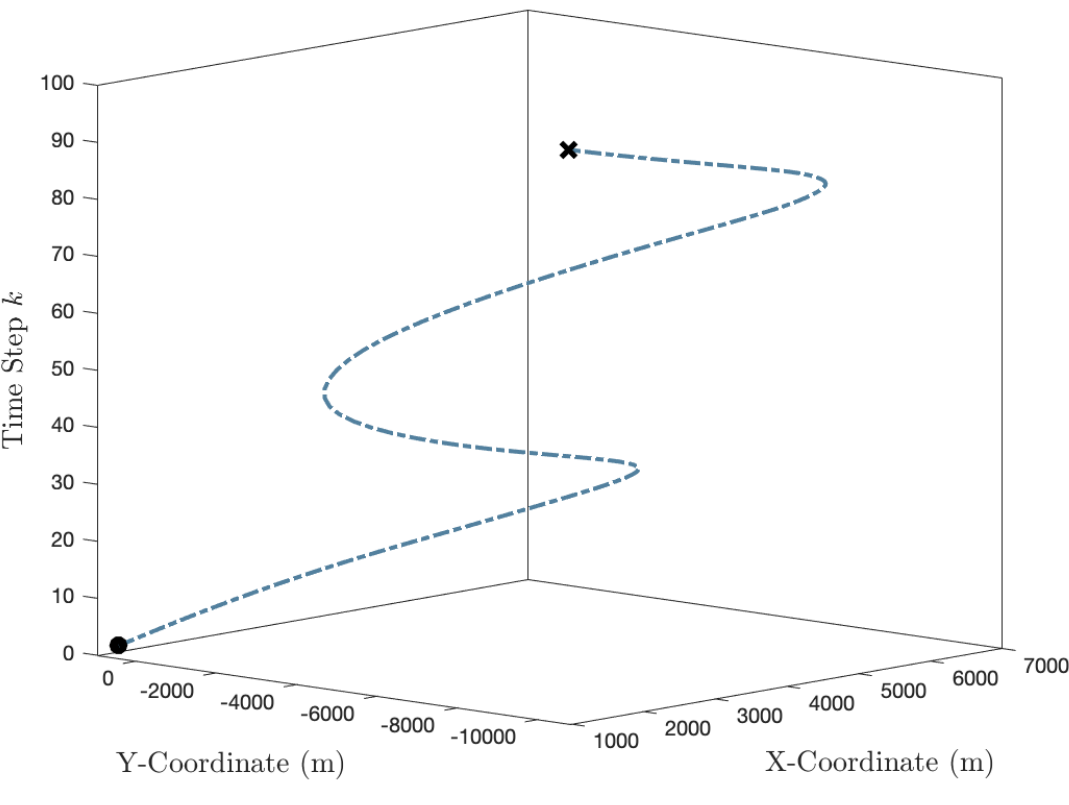}
	\caption{Scenario 2: A trajectory of an object maneuvering in two-dimensional Cartesian space with time shown vertically, where $\sbullet[1.2]$ and $\scross[0.8]$ indicate the initial and end points, respectively.}
	\label{fig:tracjectory_3D}
\end{figure}

The simulation parameter settings for the dual-tracking system are specified in Table~\ref{Table:sim_1}.
The object maneuvers in a two-dimensional trajectory with a duration of $100$ time steps
as shown in Fig.\ref{fig:tracjectory_3D}. The object's state and covariance are initialized as:
\begin{align}
\bx_{0}&=[1000~\mathrm{m}, 300~\mathrm{m/s}, 1000~\mathrm{m}, 0~\mathrm{m/s},  -3^{\circ}/\mathrm{s}]^{T} , \\
\bP_{0}&=\text{diag}[100~\mathrm{m^2}, 10~\mathrm{m^2/s^2}, 100~\mathrm{m^2}, 10~\mathrm{m^2/s^2}, \nonumber\\
&\hspace{4.3cm} 100\times10^{-3}~\mathrm{rad^2/s^2}] . 
\end{align}
The performance results are evaluated by averaging $10,000$ \ac{mc} simulation runs. The object trajectory and parameter settings are identical to the simulation realization in~\cite{Alotaibi-2024-TL_UKF,Alotaibi-2024_TL_CKF_arXiv} to present a fair comparison with the previous results obtained
from incorporating \ac{btlf} into the \ac{ukf} and \ac{ckf} local approximation approaches. 

\input{Table_1}

\subsubsection{Performance Evaluation}

\begin{figure}
	\centering
	\includegraphics[trim={2cm 6.8cm 2cm 7cm},clip, scale=0.7]{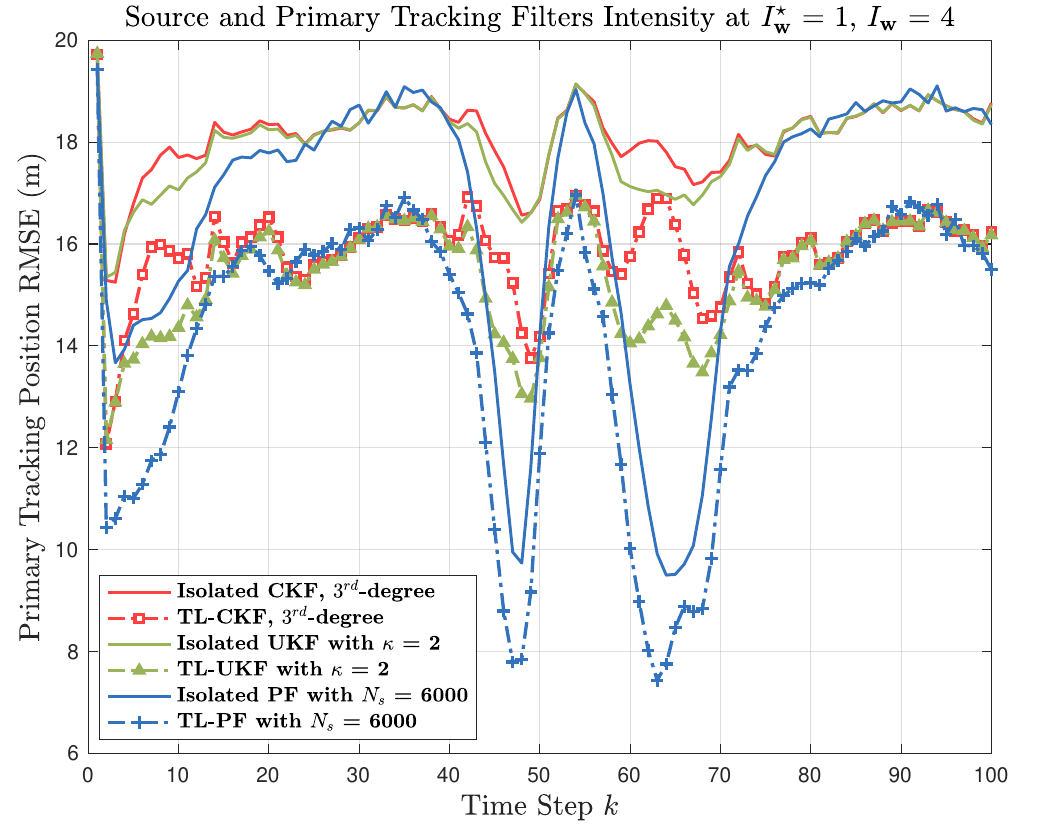}
	\caption{Scenario 2: A comparison of obtained \ac{rmse} performance plots versus time step for the proposed algorithm \ac{tl-pf} (blue dashed line with $\sadd$ marks), \ac{tl-ukf}~\cite{Alotaibi-2024_TL_CKF_arXiv}, \ac{tl-ckf}~\cite{Alotaibi-2024_TL_CKF_arXiv}, and isolated filters.} 
	\label{fig:complex_PF_RMSE_vs_Time_6000}
\end{figure}

Similarly, we evaluate the proposed \ac{tl-pf} using the tracking Scenario~2 in Fig.\ref{fig:tracjectory_3D}. The performance metric of interest is the \ac{rmse} of the object's position $(x_{k}, y_{k})$ for the primary tracking filter. The performance of the proposed \ac{tl-pf} and isolated \ac{pf} are evaluated in terms of \ac{rmse} per time step provided in Fig.~\ref{fig:complex_PF_RMSE_vs_Time_6000} in addition to isolated and incorporating transfer learning of local approaches, \ac{ukf} and third-degree \ac{ckf}, previously presented in \cite{Alotaibi-2024_TL_CKF_arXiv}. The proposed \ac{tl-pf} algorithm with $N_s = 6000$ particles (blue dashed line with $\sadd$ marks) achieves \ac{rmse} values of $7.85$~m and $14.38$~m at time steps $k=48$~s and $k=75$~s, respectively, under the noise intensities of $I^{\TR}_{\bw} = 4$ (primary sensor) and $I^{\LR}_{\bw} = 1$ (source sensor). In comparison, the isolated version of \ac{pf} (blue solid line) has \ac{rmse} values of $9.73$~m and $17.34$~m at the same noise level and time steps. Furthermore, the proposed \ac{tl-pf} algorithm (with $N_s = 6000$) achieves an \ac{rmse} of $13.10$~m at time step $k=10$~s, outperforming the third-degree \ac{tl-ckf} and \ac{tl-ukf} (with $\ka=2$), which obtain \ac{rmse} values of $15.72$~m and $14.35$~m, respectively, under the same conditions. From Fig.~\ref{fig:complex_PF_RMSE_vs_Time_6000}, we see that the performance of the proposed \ac{tl-pf} is always superior to that of isolated approaches. For most of the trajectory, \ac{tl-pf} performs significantly better than \ac{tl-ckf} and \ac{tl-ukf}. In the intervals of $(20, 40)$~s and $(85, 95)$~s, \ac{tl-pf} performs comparably to \ac{tl-ckf} and \ac{tl-ukf}. However, we note that the performance of \ac{tl-pf} can be further improved by increasing the number of particles, $N_s$, at the expense of higher computational cost. 

Furthermore, the \ac{rmse} curves of the algorithms presented in Fig.~\ref{fig:complex_PF_RMSE_vs_Time_6000} have fluctuations in the interval of $(40, 75)$~s. 
We compare the performance results from Scenario~1 (Fig.~\ref{fig:Linear_PF_RMSE_vs_Time_6000} ) with those from Scenario~2 (Fig.\ref{fig:complex_PF_RMSE_vs_Time_6000}). We note that the fluctuations in the interval of $(40, 75)$~s are due to the nature of the maneuvering trajectory where the object track has high maneuvering turns in the same time interval, as obviously shown in Fig.~\ref{fig:tracjectory_3D}. During this time interval, the proposed \ac{tl-pf} demonstrates significant strength in capturing object maneuvering turns and estimates the object's position more accurately with lower \ac{rmse} values compared to other plotted algorithms. For example, the \ac{tl-pf} algorithm (with $N_s = 6000$) achieves an \ac{rmse} value of $7.79$~m during the high maneuvering turns interval at time step $k=47$~s, while the \ac{tl-ukf} (green dashed line) has an \ac{rmse} value of $13.75$~m. 

\subsubsection{Overall Performance Gain}

\begin{figure}
	\centering
	\includegraphics[trim={2cm 6.8cm 2cm 7cm},clip, scale=0.7]{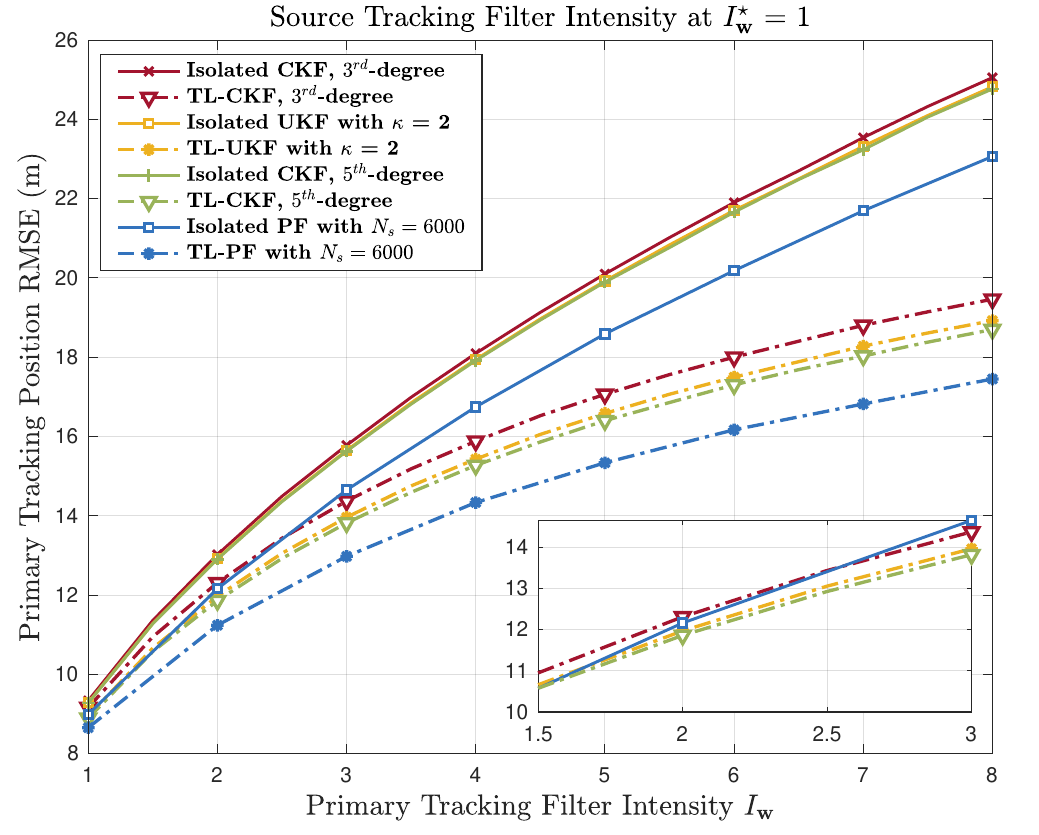}
	\caption{Scenario 2: The overall \ac{rmse} performance of the proposed \ac{tl-pf} algorithm (with $N_s =6000$) under varying primary noise intensity $I^{\TR}_{\bw}$, plotted alongside the performance of local \ac{btlf} approaches in~\cite{Alotaibi-2024_TL_CKF_arXiv}.}
	\label{fig:complex_PF_RMSE_vs_IW_6000}
\end{figure}

To further demonstrate the behavior of the proposed \ac{tl-pf} algorithm, we evaluate its
overall \ac{rmse} performance, averaged over $100$ time steps and $10,000$ \ac{mc} iterations, under various levels of primary noise intensity $I^{\TR}_{\bw}$, while fixing the intensity level of source noise $I^{\LR}_{\bw}$ as follows:
\begin{align}
\text{Overall RMSE} = \frac{1}{K} \sum_{k=1}^{K} \sqrt{\frac{1}{MC} \sum_{m=1}^{MC} (\text{True\ Pos.}_{m,k} - \text{Est.\ Pos.}_{m,k})^2} .
\end{align}
As shown in Fig.~\ref{fig:complex_PF_RMSE_vs_IW_6000}, the accuracy gain of the \ac{tl-pf} (blue dashed line marked with~$\sstar$) compared to the isolated traditional \ac{pf} increases as the level of primary noise intensity $I^{\TR}_{\bw}$ increases. For instance, under $I^{\TR}_{\bw} =4$, the \ac{tl-pf} (with $N_s = 6000$) achieves an \ac{rmse} value of $14.33$~m, while the isolated \ac{pf} has an \ac{rmse} of $16.74$~m with a reduction in estimation error of approximately $2.41$~m. Similarly, under $I^{\TR}_{\bw} =8$, the \ac{tl-pf} has an \ac{rmse} of $17.45$~m compared to $23.06$~m for the isolated \ac{pf} with a reduction of $5.61$~m. Furthermore, the \ac{tl-pf} (with $N_s = 6000$) outperforms the fifth-degree \ac{tl-ckf} (green dashed line with $\stria$~marks) and is capable of achieving an accuracy gain value of approximately $1.24$~m under a level of primary noise intensity $I^{\TR}_{\bw} = 8$.

\begin{figure}
	\centering
	\includegraphics[trim={2cm 6.8cm 2cm 7cm},clip, scale=0.7]{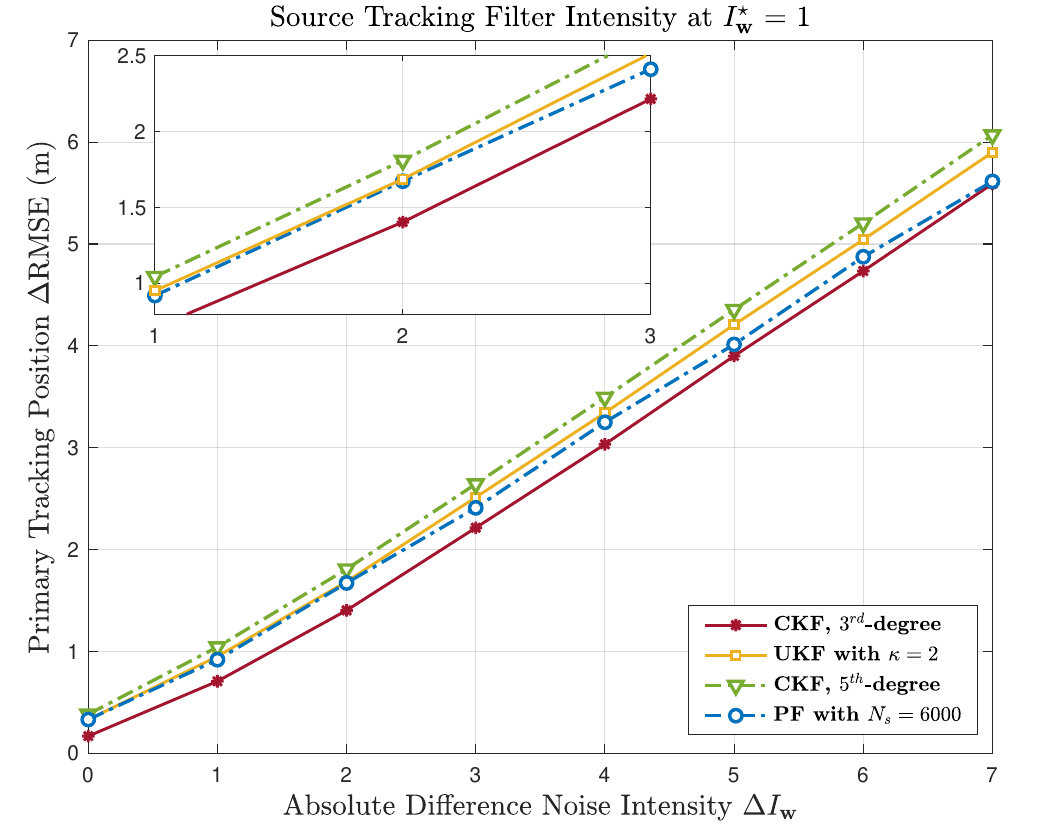}
	\caption{Scenario 2: The \ac{rmse} gain $\Delta \text{RMSE}$ of incorporating \ac{btl} for third-degree \ac{ckf} (maroon solid line), \ac{ukf} (yellow solid line), fifth-degree \ac{ckf} (green dashed line), and \ac{pf} (blue dashed line) versus the absolute difference between the primary and source noise intensities $\Delta I_{\bw}$.}
	\label{fig:complex_Delta_RMSE_vs_Delta_IW}
\end{figure}

To investigate the performance gain of incorporating \ac{btlf} into several tracking algorithms, we evaluate the primary tracking position \ac{rmse} gain of using \ac{btlf} compared to isolated filters calculated as
\be
\Delta \text{RMSE} = \text{RMSE}^{\text{Isolated}} - \text{RMSE}^{\text{TL}}  ,
\ee
under varying $\Delta I_{\bw}$, which is the absolute value of the difference between the primary and source noise intensities computed as 
\be
\Delta I_{\bw} =  |I^{\TR}_{\bw} - I^{\LR}_{\bw}| .
\ee
As shown in Fig.~\ref{fig:complex_Delta_RMSE_vs_Delta_IW}, the $\Delta \text{RMSE}$ for third-degree \ac{ckf}, \ac{ukf}, fifth-degree \ac{ckf}, and \ac{pf} (with $N_s = 6000$) increases as the absolute difference between the primary and source noise intensities $\Delta I_{\bw}$ increases. This describes the proportional relationship between $\Delta \text{RMSE}$ and $\Delta I_{\bw}$ where $\Delta \text{RMSE}$ increases almost linearly. For instance, at $\Delta I_{\bw} = 3 $, the \ac{pf} (with $N_s = 6000$) and third-degree \ac{ckf} achieve $\Delta \text{RMSE}$ values of $2.41$~m and $2.21$~m, respectively. However, the \ac{pf} (with $N_s = 6000$) has $\Delta \text{RMSE}$ value of approximately $5.61$~m, which is less than $\Delta \text{RMSE}$ value of $6.06$~m for the fifth-degree \ac{ckf} at $\Delta I_{\bw} = 7 $. We note that the fifth-degree \ac{ckf} performs significantly better in terms of the primary tracking position \ac{rmse} gain $\Delta \text{RMSE}$ for most of the absolute difference values between the primary and source noise intensities $\Delta I_{\bw}$, as illustrated in Fig.~\ref{fig:complex_Delta_RMSE_vs_Delta_IW}. These $\Delta \text{RMSE}$ curves can serve as a guideline for whether \ac{btlf} can be incorporated in a certain system depending on the difference between the intensities of the noise.

\subsubsection{Varying Particles Number and Computational Time}

\begin{figure}
	\centering
	\includegraphics[trim={2.1cm 6.8cm 1.6cm 7cm},clip, scale=0.7]{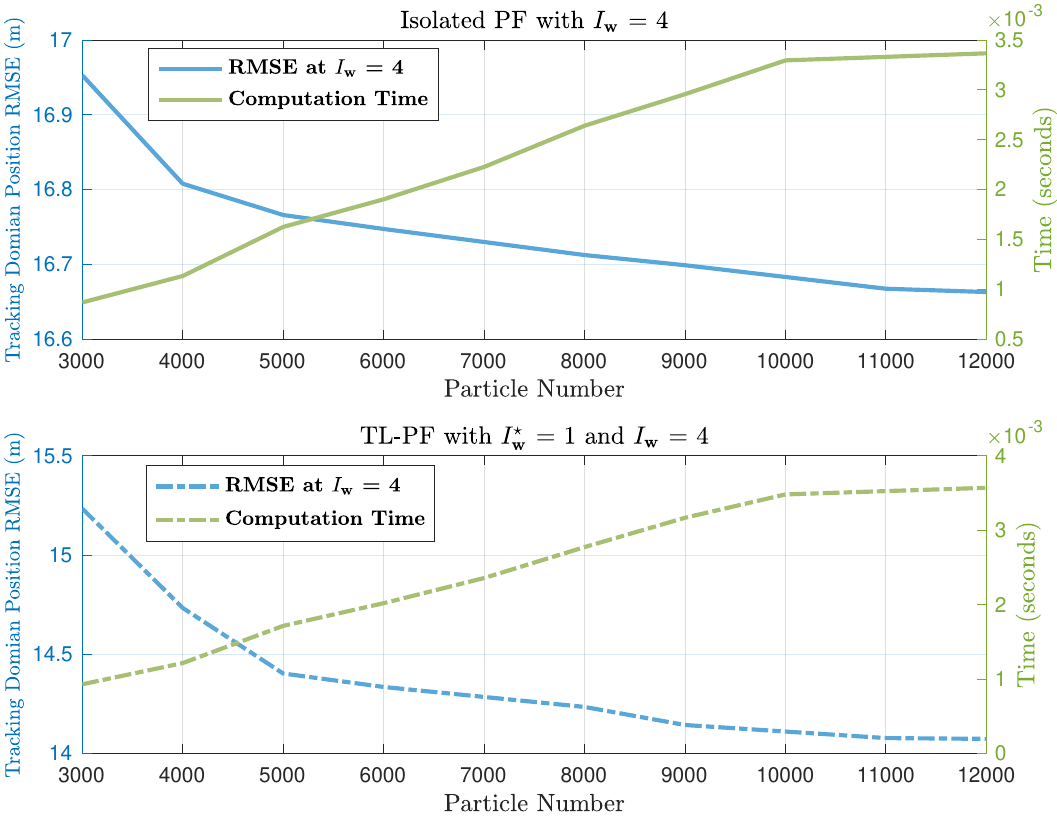}%[trim={2.1cm 6.8cm 1.6cm 7cm},clip, scale=0.485]
	\caption{Scenario 2: \ac{rmse} (blue) and the computation time (green) curves of isolated (top) and incorporating transfer learning (bottom) for \ac{pf} at primary noise intensity $I^{\TR}_{\bw} = 4$.}
	\label{fig:complex_Comp_Time_&_RMSE_vs_PF_Num}
\end{figure}

To investigate the effect of particles number on the accuracy performance of the proposed \ac{tl-pf} algorithm, we performed a simulation analysis with varying number of particles $N_s = 3000\rightarrow12000$ under noise intensity levels at $I^{\LR}_{\bw} = 1$ and $I^{\TR}_{\bw} = 4$. The obtained results, shown in Fig.~\ref{fig:complex_Comp_Time_&_RMSE_vs_PF_Num}, represent that increasing the number of particles $N_s$ leads to higher accuracy performance (lower value of \ac{rmse}) for both the \ac{tl-pf} and isolated \ac{pf} algorithms. This describes the inverse proportional relationship between the \ac{rmse} and the number of particles. Furthermore, the computational time per time step $k$, evaluated on a laptop with a $3.1$~GHz Core i$7$ processor, of both the \ac{tl-pf} (bottom plot with green dashed line) and isolated \ac{pf} (top plot with green solid line) is proportional to the number utilized. Moreover, the sensitivity level of the proposed \ac{tl-pf} algorithm to the number of particles is greater than the isolated version of \ac{pf}. For instance, with $N_s = 3000$ particles, the \ac{tl-pf} achieves \ac{rmse} of $15.23$~m, while the isolated \ac{pf} has an \ac{rmse} of $16.95$~m. By increasing the number of particles to $N_s = 12000$, the \ac{rmse} of \ac{tl-pf} improves to $14.07$~m, while the \ac{rmse} of isolated \ac{pf} improves only to $16.66$~m. As the number of particles increases from $3000$ to $12000$, the \ac{tl-pf} gains an accuracy of $1.16$~m compared to $0.29$~m for the isolated \ac{pf}. 
%To evaluate the computational time performance of tracking algorithms, simulation realizations were performed on a laptop with a $3.1$~GHz Core i$7$ processor. 
Table~\ref{Table:sim_2} presents a comparative analysis of local and global approaches integrated with \ac{btlf} and their corresponding isolated filters in terms of \ac{rmse} and computational process time per time step. Note that the \ac{rmse} results for the first three rows in Table~\ref{Table:sim_2}, which represent local approaches, have previously been obtained in~\cite{Alotaibi-2024_TL_CKF_arXiv}. These results are included in this article to provide a comparative baseline for the proposed algorithm using the global approach. Table~\ref{Table:sim_2} shows that integrating of the global approach, \ac{pf}, with \ac{btlf} achieves a superior reduction in \ac{rmse} results compared to \ac{rmse} performances of \ac{tl-ukf}, third-degree \ac{tl-ckf}, and fifth-degree \ac{tl-ckf}. 
This \ac{rmse} improvement comes with a modest increase in processing time, from $0.296$~ms (fifth-degree \ac{tl-ckf}) to $0.931$~ms 
(\ac{tl-pf} with $N_s = 3000$), as shown in Table~\ref{Table:sim_2}.

\input{Table_2_v2}

%% file: Table_1.tex
\begin{table}[htbp]
	\caption{Scenario 2: Simulation Parameter Settings}
	\label{Table:sim_1}
	\centering

\setlength{\tabcolsep}{32.65 pt} % Default value: 6pt
\renewcommand{\arraystretch}{2.2} % Default value: 1

\begin{tabularx}{\columnwidth}{cc|cc}	
        \toprule

		Parameter & Value & Parameter & Value\\
		\cmidrule{1-4}
		\rowcolor{gray!15} 
            $n_\bx$ & $5$ &
		$n_\bz$ & $2$ \\
		$T_s$ & $1\ \mathrm{s}$ &
            $K$ & $100$ \\
		\rowcolor{gray!15} 
            MC~runs & $10,000$ &
            $\kappa$ & $2$ \\
		$q_1$ & $0.1\ \mathrm{m^2/s^4}$ &
		$q_2$ & $1.75\times10^{-2}\ \mathrm{rad^2/s^3}$ \\
		\rowcolor{gray!15} 
            $\sigma_{r}$ & $10\ \mathrm{m}$ &
		$\sigma_{\zeta}$ & $\sqrt{10}\times 10^{-3}\ \mathrm{rad}$ \\
		 $I^{\LR}_{\bw}$ & $1$ &
		 $I^{\TR}_{\bw}$ & $1\rightarrow8$\\
   
        \bottomrule

\end{tabularx}

\end{table}

%% file: Table_2_v2.tex
\begin{table*}[htbp]%[tbp]
	\caption{Scenario 2: Position RMSE and Computation Time for Primary Tracking Filters with $I^{\TR}_{\bw} = 4$}
        \centering
	\label{Table:sim_2}

\setlength{\tabcolsep}{10.95pt} % Default value: 6pt
\renewcommand{\arraystretch}{2.2} % Default value: 1

%\begin{tabular*}{\columnwidth}{@{\extracolsep{\fill}} lc cc|cc}	
%\begin{tabularx}{\textwidth}{@{}llcc|cc@{}}
\begin{tabularx}{\textwidth}{l|c| cc|cc}	
%\begin{tabularx}{\textwidth}{LY YY|YY}	
%\begin{tabular}{|L{40pt} C{30pt} C{37pt} C{30pt} C{37pt} C{30pt}|}
        \toprule
        
	 \multirow{2}{*}{Filter} & \multirow{2}{*}{Points} & \multicolumn{2}{c|}{Isolated} & %
	\multicolumn{2}{c}{BTLF}\\
	%\cline{2-7}
	%& \multicolumn{2}{c|}{Value} & \multicolumn{2}{c|}{Value} & \\
	%\cline{2-7}

	 \cmidrule{3-4}  \cmidrule{5-6}
  
     &  & RMSE ($\mathrm{m}$) & Time per Step ($\mathrm{ms}$) & RMSE ($\mathrm{m}$) & Time per Step ($\mathrm{ms}$) \\
	%\hline
        \cmidrule{1-6}
        \rowcolor{gray!15} 
        %\makebox[-2.7pt][c]{\fboxsep2.5pt\colorbox[gray]{0.9}{\strut\hspace*{0.98\columnwidth}}}
	  CKF, $3^{rd}$-degree & 10 
                             & 18.0973 & 0.154 
                             & 15.8815 & 0.188\\
                             
	UKF, $\ka = 2$ & 11  
                             & 17.9451   & 0.165
                             & 15.4299   & 0.206 \\
                             
         \rowcolor{gray!15}                     
	%\makebox[-2.7pt][l]{\fboxsep2.5pt\colorbox[gray]{0.9}{\strut\hspace*{0.98\columnwidth}}}
        CKF, $5^{th}$-degree & 51  
                             & 17.9178  & 0.225
                             & 15.2728  & 0.297 \\

        SIR-PF               & 3000  
                             & 16.9530  & 0.867 
                             & 15.2341  & 0.931 \\
                             
         \rowcolor{gray!15}                     
	%\makebox[-2.7pt][l]{\fboxsep2.5pt\colorbox[gray]{0.9}{\strut\hspace*{0.98\columnwidth}}}
        SIR-PF               & 6000  
                             & 16.7473  & 1.902 
                             & 14.3359  & 2.020 \\

        SIR-PF               & 9000  
                             & 16.6988  & 2.957 
                             & 14.1443  & 3.167 \\
                             
	%\makebox[-2.7pt][l]{\fboxsep2.5pt\colorbox[gray]{0.9}{\strut\hspace*{0.98\columnwidth}}}
        \rowcolor{gray!15} 
        SIR-PF               & 12000   
                             & 16.6630  & 3.367 
                             & 14.0746  & 3.569 \\

 \bottomrule

\end{tabularx}

\end{table*}

%% file: Conclusion.tex
\section{Conclusion}
\label{Sec:conclusion}

We developed a \ac{btl} approach for particle filtering, \ac{tl-pf}, in a dual-tracking system with asymmetric measurement noise intensities at the primary and source sensors. By leveraging predicted measurement information transferred from the source sensor, the primary sensor, which experiences the higher noise intensity, is able to achieve a significant gain in tracking performance compared to an isolated particle filter, without applying transfer learning. We also compared the performance of the proposed transfer learning particle filter schemes with earlier schemes based on applying Bayesian transfer learning to the unscented Kalman filter and the cubature Kalman filter, referred to as \ac{tl-ukf} and \ac{tl-ckf}, respectively. Furthermore, the \ac{tl-pf} shows greater sensitivity to the number of particles compared to the isolated particle filter. Although increasing the number of particles improves tracking performance, it also increases computational time, which can be a challenge in real-time processing scenarios. However, a moderate number of particles can be effectively utilized in certain applications, particularly with the advancements in hardware technology that allow for greater computational efficiency. For the number of particles $N_s = 6000$, the process times per step are approximately $2$~ms and $1.9$~ms for the \ac{tl-pf} and the isolated particle filter, respectively. Furthermore , we observe from the numerical results that the relative gain of the \ac{tl-pf} is linearly proportional to the absolute difference value between the source and primary sensors. 

For future work, it would be interesting to explore a hybrid system that incorporates a decision step at either the source or the primary sensor to determine whether to perform transfer learning at each time step. In addition, the framework can be generalized to involve more than a single source tracking system by transferring predicted observations and feeding them into the primary tracking system using the proposed \ac{tl-pf}. Moreover, investigating hardware acceleration approaches to reduce the computational time of \ac{tl-pf} is 
another promising direction for future work.

%%% 
% 1. I need to add a sentence about increasing the number of particles and sensitivity of TL-PF compared to isolated PF.
% 2. I need to add a sentence about the computation time.
% 3. I need to add a future work about adding a decsion step to make the system hybrid. 

%% file: Proof_TL_PF.tex
\section{Particle Filter with Bayesian Transfer Learning Derivation}
\label{TL_PF_Derivation_Section}

\subsection{Source Tracking Filter}

The object state posterior density in~\eqref{eq:BTL_sr_state_2} is given by
\be\label{eq:Der:BTL_sr_state_2}
p(\bx^{\LR}_{0:k} \mid \bz^{\LR}_{1:k}) \approx \sum_{i=1}^{N_s} \ w_{k}^{\LR (i)} \delta \left( \bx^{\LR}_{0:k} - \bx^{\LR (i)}_{0:k}\right) ,
\ee
and associated particle weights are computed as
\be\label{eq:Der:BTL_sr_weigths_1}
w_{k}^{\LR (i)} \propto \frac{p(\bx^{\LR (i)}_{0:k} \mid \bz^{\LR}_{1:k})}{q(\bx^{\LR (i)}_{0:k} \mid \bz^{\LR}_{1:k})} .
\ee
The numerator in~\eqref{eq:Der:BTL_sr_weigths_1} can be written as 
\begin{align}
p(\bx^{\LR (i)}_{0:k} \mid \bz^{\LR}_{1:k}&) = p( \bx^{\LR (i)}_{0:k} \mid \bz^{\LR}_{k}, \bz^{\LR}_{1:k-1})\nonumber\\
&= \frac{p(\bz^{\LR}_{k} \mid \bx^{\LR (i)}_{0:k}, \bz^{\LR}_{1:k-1}) \ p(\bx^{\LR (i)}_{0:k} \mid \bz^{\LR}_{1:k-1})}{p(\bz^{\LR}_{k} \mid \bz^{\LR}_{1:k-1})} , \label{eq:Der:BTL_sr_weigths_2}
\end{align}
where $p(\bz^{\LR}_{k} \mid \bx^{\LR (i)}_{0:k}, \bz^{\LR}_{1:k-1}) = p(\bz^{\LR}_{k} \mid \bx^{\LR (i)}_{k})$ is due to the fact that the measurement $\bz^{\LR}_{k}$ given the current state $\bx^{\LR (i)}_{k}$ is conditionally independent of all previous states $\bx^{\LR (i)}_{0:k-1}$ and measurements $\bz^{\LR}_{1:k-1}$, and $p(\bx^{\LR (i)}_{0:k} \mid \bz^{\LR}_{1:k-1}) = p(\bx^{\LR (i)}_{k} \mid \bx^{\LR (i)}_{k-1}) \ p(\bx^{\LR (i)}_{0:k-1} \mid \bz^{\LR}_{1:k-1})$ is derived according to the first-order Markov chain assumption. Furthermore, the density $p(\bx^{\LR (i)}_{0:k} \mid \bz^{\LR}_{1:k})$ in~\eqref{eq:Der:BTL_sr_weigths_2} can be rewritten as 
\begin{align}\label{eq:Der:BTL_sr_weigths_3}
p(\bx^{\LR (i)}_{0:k} \mid \bz^{\LR}_{1:k}) \propto p(\bz^{\LR}_{k} \mid \bx^{\LR (i)}_{k}) \ p(\bx^{\LR (i)}_{k} \mid \bx^{\LR (i)}_{k-1}) \ p(\bx^{\LR (i)}_{0:k-1} \mid \bz^{\LR}_{1:k-1}) .
\end{align}
The denominator in~\eqref{eq:Der:BTL_sr_weigths_1} can be factorized similarly to~\cite{ArulaMGC-2002-Tutorial_PF_EKF} as
\begin{align}
q(\bx^{\LR (i)}_{0:k} \mid \bz^{\LR}_{1:k}) = q(\bx^{\LR (i)}_{k} \mid \bx^{\LR (i)}_{0:k-1}, \bz^{\LR}_{1:k}) \ q(\bx^{\LR (i)}_{0:k-1} \mid \bz^{\LR}_{1:k-1}) \label{eq:Der:BTL_sr_weigths_4},
\end{align}
where $q(\bx^{\LR (i)}_{k} \mid \bx^{\LR (i)}_{0:k-1}, \bz^{\LR}_{1:k}) = q(\bx^{\LR (i)}_{k} \mid \bx^{\LR (i)}_{k-1}, \bz^{\LR}_{1:k})$ due to the dependency only on $\bx^{\LR (i)}_{k-1}$ and conditionally independent of all previous states $\bx^{\LR (i)}_{0:k-2}$. Therefore, the particle weights can be rewritten by substituting~\eqref{eq:Der:BTL_sr_weigths_3} and~\eqref{eq:Der:BTL_sr_weigths_4} into~\eqref{eq:Der:BTL_sr_weigths_1} as
\be\label{eq:Der:BTL_sr_weigths_5}
w_{k}^{\LR (i)} \propto w_{k-1}^{\LR (i)}\ \frac{p(\bz^{\LR}_{k} \mid \bx^{\LR (i)}_{k}) \ p(\bx^{\LR (i)}_{k} \mid \bx^{\LR (i)}_{k-1})}{q(\bx^{\LR (i)}_{k} \mid \bx^{\LR (i)}_{k-1}, \bz^{\LR}_{1:k})} .
\ee

The predicted observation posterior density in~\eqref{eq:BTL_sr_pred_obser_1} is approximated by weighted particles as
\be\label{eq:Der:BTL_sr_pred_obser_1}
p(\bet^{\LR}_{1:k+1} \mid \bz^{\LR}_{1:k}) \approx \sum_{i=1}^{N_s} \ w_{k+1}^{\LR \bet (i)} \delta \left( \bet^{\LR}_{1:k+1} - \bet^{\LR (i)}_{1:k+1}\right) .
\ee
The weights $w_{k+1}^{\LR \bet (i)}$ in~\eqref{eq:Der:BTL_sr_pred_obser_1} are computed as 
\be\label{eq:Der:BTL_sr_pred_obser_2}
w_{k+1}^{\LR \bet (i)} \propto \frac{p(\bet^{\LR (i)}_{1:k+1} \mid \bz^{\LR}_{1:k})}{q(\bet^{\LR (i)}_{1:k+1} \mid \bz^{\LR}_{1:k})} ,
\ee
where the numerator $p(\bet^{\LR (i)}_{1:k+1} \mid \bz^{\LR}_{1:k})$ can be written as
\begin{align}\label{eq:Der:BTL_sr_pred_obser_3}
&p(\bet^{\LR (i)}_{1:k+1} \mid \bz^{\LR}_{1:k}) = p(\bet^{\LR (i)}_{k+1} , \bet^{\LR (i)}_{1:k} \mid \bz^{\LR}_{1:k}) \nonumber \\
&= p(\bet^{\LR (i)}_{k+1} \mid \bx^{\LR (i)}_{0:k+1}, \bz^{\LR}_{1:k} , \bet^{\LR (i)}_{1:k})\ p(\bx^{\LR (i)}_{0:k+1} \mid \bz^{\LR}_{1:k} , \bet^{\LR (i)}_{1:k}) . 
\end{align}
Similarly to deriving particle weights of the object state density, the predicted observation $\bet^{\LR (i)}_{k+1}$ depends only on the predicted state $\bx^{\LR (i)}_{k+1}$ and is conditionally independent of all previous states $\bx^{\LR (i)}_{0:k}$ and measurements $\bz^{\LR}_{1:k}$ resulting in $p(\bet^{\LR (i)}_{k+1} \mid \bx^{\LR (i)}_{0:k+1}, \bz^{\LR}_{1:k} , \bet^{\LR (i)}_{1:k}) = p(\bet^{\LR (i)}_{k+1} \mid \bx^{\LR (i)}_{k+1})$. Furthermore, the second term $p(\bx^{\LR (i)}_{0:k+1} \mid \bz^{\LR}_{1:k} , \bet^{\LR (i)}_{1:k})$ in~\eqref{eq:Der:BTL_sr_pred_obser_3} can be written as
\begin{align}
p(\bx^{\LR (i)}_{0:k+1} \mid \bz^{\LR}_{1:k} , \bet^{\LR (i)}_{1:k}) &= p(\bx^{\LR (i)}_{k+1} \mid \bx^{\LR (i)}_{0:k}, \bz^{\LR}_{1:k} , \bet^{\LR (i)}_{1:k})\ p(\bx^{\LR (i)}_{0:k} \mid \bz^{\LR}_{1:k} , \bet^{\LR (i)}_{1:k}) \label{eq:Der:BTL_sr_pred_obser_4}\\
& = p(\bx^{\LR (i)}_{k+1} \mid \bx^{\LR (i)}_{k})\ p(\bx^{\LR (i)}_{0:k} \mid \bz^{\LR}_{1:k} ) \label{eq:Der:BTL_sr_pred_obser_5} ,
\end{align}
where simplification from~\eqref{eq:Der:BTL_sr_pred_obser_4} to~\eqref{eq:Der:BTL_sr_pred_obser_5} due to the first-order Markov chain assumption and to conditionally independent of self-predicted observations $\bet^{\LR (i)}_{1:k}$. The denominator in~\eqref{eq:Der:BTL_sr_pred_obser_2} can be factorized as
\begin{align}\label{eq:Der:BTL_sr_pred_obser_6}
q(\bet^{\LR (i)}_{1:k+1} \mid \bz^{\LR}_{1:k}) = q(\bet^{\LR (i)}_{k+1} \mid \bx^{\LR (i)}_{0:k}, \bz^{\LR}_{1:k}, \bet^{\LR (i)}_{1:k}) \ q(\bx^{\LR (i)}_{0:k} \mid \bz^{\LR}_{1:k}, \bet^{\LR (i)}_{1:k}) , 
\end{align}
where $q(\bx^{\LR (i)}_{0:k} \mid \bz^{\LR}_{1:k}, \bet^{\LR (i)}_{1:k}) = q(\bx^{\LR (i)}_{0:k} \mid \bz^{\LR}_{1:k})$ is due to the conditionally independent of the self-predicted observations. Moreover, the predicted observation $\bet^{\LR (i)}_{k+1}$ given the current state $\bx^{\LR (i)}_{k}$ is conditionally independent of all previous states $\bx^{\LR (i)}_{0:k-1}$ and predicted observations $\bet^{\LR (i)}_{1:k}$ yielding to
\be \label{eq:Der:BTL_sr_pred_obser_7}
q(\bet^{\LR (i)}_{k+1} \mid \bx^{\LR (i)}_{0:k}, \bz^{\LR}_{1:k}, \bet^{\LR (i)}_{1:k}) = q(\bet^{\LR (i)}_{k+1} \mid \bx^{\LR (i)}_{k}, \bz^{\LR}_{1:k}) . 
\ee
By substituting derived and simplified densities from~\eqref{eq:Der:BTL_sr_pred_obser_3} to~\eqref{eq:Der:BTL_sr_pred_obser_7} into~\eqref{eq:Der:BTL_sr_pred_obser_2}, the weights $w_{k+1}^{\LR \bet (i)}$ can be rewritten as 
\be \label{eq:Der:BTL_sr_pred_obser_8}
w_{k+1}^{\LR \bet (i)} \propto w_{k}^{\LR (i)}\ \frac{p(\bet^{\LR (i)}_{k+1} \mid \bx^{\LR (i)}_{k+1})\ p(\bx^{\LR (i)}_{k+1} \mid \bx^{\LR (i)}_{k})}{q(\bet^{\LR (i)}_{k+1} \mid \bx^{\LR (i)}_{k}, \bz^{\LR}_{1:k}) } .
\ee

\subsection{Primary Tracking Filter}

The overall posterior density that leverages transferred knowledge from the source tracking filter in~\eqref{eq:BTL_tr_state_2} is approximately computed via weighted particles as 
\be\label{eq:Der:BTL_tr_state_2}
p(\bx^{\TR}_{k} \mid \bz^{\TR}_{1:k}, \bet^{\LR}_{2:k}) \approx \sum_{i=1}^{N_s} \ w_{k}^{\TR (i)} \delta \left( \bx^{\TR}_{k} - \bx^{\TR (i)}_{k}\right) ,
\ee
and associated weights $w_{k}^{\TR (i)}$ are defined as 
\be\label{eq:Der:BTL_tr_weigths_1}
w_{k}^{\TR (i)} \propto \frac{p(\bx^{\TR (i)}_{0:k} \mid \bz^{\TR}_{1:k}, \bet^{\LR}_{2:k})}{q(\bx^{\TR (i)}_{0:k} \mid \bz^{\TR}_{1:k}, \bet^{\LR}_{2:k})} .
\ee
The numerator in~\eqref{eq:Der:BTL_tr_weigths_1} can be factorized for derivation as $p(\bx^{\TR (i)}_{0:k} \mid \bz^{\TR}_{1:k}, \bet^{\LR}_{2:k}) = p(\bx^{\TR (i)}_{0:k} \mid \bz^{\TR}_{k}, \bz^{\TR}_{1:k-1}, \bet^{\LR}_{k}, \bet^{\LR}_{2:k-1})$ and written as 
\begin{align}
&\hspace{-.1cm} p(\bx^{\TR (i)}_{0:k} \mid \bz^{\TR}_{k}, \bz^{\TR}_{1:k-1}, \bet^{\LR}_{k}, \bet^{\LR}_{2:k-1})=  \nonumber\\
&\hspace{2cm} \frac{p(\bz^{\TR}_{k}\! \mid\! \bx^{\TR (i)}_{0:k},\! \bz^{\TR}_{1:k-1},\! \bet^{\LR}_{k},\! \bet^{\LR}_{2:k-1}\!)\ p(\bet^{\LR}_{k}\! \mid\! \bx^{\TR (i)}_{0:k},\! \bz^{\TR}_{1:k-1},\! \bet^{\LR}_{2:k-1}\!)}{p(\bz^{\TR}_{k}\! \mid\! \bz^{\TR}_{1:k-1}) \ p(\bet^{\LR}_{k}\! \mid\! \bet^{\LR}_{2:k-1})} .\ p(\bx^{\TR (i)}_{0:k} \mid \bz^{\TR}_{1:k-1}, \bet^{\LR}_{2:k-1}) ,  \label{eq:Der:BTL_tr_weigths_2}
\end{align}
where $p(\bz^{\TR}_{k} \mid \bx^{\TR (i)}_{0:k}, \bz^{\TR}_{1:k-1}, \bet^{\LR}_{k}, \bet^{\LR}_{2:k-1}) = p(\bz^{\TR}_{k} \mid \bx^{\TR (i)}_{k})$ and $p(\bet^{\LR}_{k} \mid \bx^{\TR (i)}_{0:k}, \bz^{\TR}_{1:k-1}, \bet^{\LR}_{2:k-1}) = p(\bet^{\LR}_{k} \mid \bx^{\TR (i)}_{k})$ are due to the conditional independency between the measurement $\bz^{\TR}_{k}$, the predicted observation $\bet^{\LR}_{k}$, all previous states $\bx^{\TR (i)}_{0:k-1}$, all previous measurements $\bz^{\TR}_{1:k-1}$, and all previous predicted observations $\bet^{\LR}_{2:k-1}$ given the current state $\bx^{\TR (i)}_{k}$. The last term on the \ac{rhs} in~\eqref{eq:Der:BTL_tr_weigths_2} can be derived on the basis of the first-order Markov chain assumption as $p(\bx^{\TR (i)}_{0:k} \mid \bz^{\TR}_{1:k-1}, \bet^{\LR}_{2:k-1}) =  p(\bx^{\TR (i)}_{k} \mid \bx^{\TR (i)}_{k-1})\ p(\bx^{\TR (i)}_{0:k-1} \mid \bz^{\TR}_{1:k-1}, \bet^{\LR}_{2:k-1})$. Therefore, the density $p(\bx^{\TR (i)}_{0:k} \mid \bz^{\TR}_{1:k}, \bet^{\LR}_{2:k})$ of the numerator in~\eqref{eq:Der:BTL_tr_weigths_1} is simplified and rewritten as 
\begin{align}
p(\bx^{\TR (i)}_{0:k} \mid \bz^{\TR}_{1:k}, \bet^{\LR}_{2:k}) \propto 
p(\bz^{\TR}_{k} \mid \bx^{\TR (i)}_{k})\ p(\bet^{\LR}_{k} \mid \bx^{\TR (i)}_{k}) p(\bx^{\TR (i)}_{k} \mid \bx^{\TR (i)}_{k-1})\ p(\bx^{\TR (i)}_{0:k-1} \mid \bz^{\TR}_{1:k-1}, \bet^{\LR}_{2:k-1}) . \label{eq:Der:BTL_tr_weigths_3}
\end{align}
Equivalently to the source tracking filter derivation, the denominator in~\eqref{eq:Der:BTL_tr_weigths_1} can be factorized as
\begin{align}
q(\bx^{\TR (i)}_{0:k} \mid \bz^{\TR}_{1:k}, \bet^{\LR}_{2:k}) = q(\bx^{\TR (i)}_{k} \mid \bx^{\TR (i)}_{0:k-1}, \bz^{\TR}_{1:k}, \bet^{\LR}_{2:k})\ q(\bx^{\TR (i)}_{0:k-1} \mid \bz^{\TR}_{1:k-1}, \bet^{\LR}_{2:k-1}) , \label{eq:Der:BTL_tr_weigths_4}
\end{align}
where the first term on the \ac{rhs} is simplified as $q(\bx^{\TR(i)}_{k} \mid \bx^{\TR (i)}_{0:k-1}, \bz^{\TR}_{1:k}, \bet^{\LR}_{2:k}) = q(\bx^{\TR (i)}_{k} \mid \bx^{\TR (i)}_{k-1}, \bz^{\TR}_{1:k}, \bet^{\LR}_{2:k})$ according to that the current state $\bx^{\TR (i)}_{k}$ given the measurement $\bz^{\TR}_{1:k}$ and the predicted observations $\bet^{\LR}_{2:k}$ depends only on the previous state $\bx^{\TR (i)}_{k-1}$ and conditionally independent of all previous states $\bx^{\TR (i)}_{0:k-2}$. By substituting derived densities of the numerator and the denominator from~\eqref{eq:Der:BTL_tr_weigths_3} and~\eqref{eq:Der:BTL_tr_weigths_4} into~\eqref{eq:Der:BTL_tr_weigths_1}, the particle weights can be rewritten as 
\be\label{eq:Der:BTL_tr_weigths_5}
w_{k}^{\TR (i)} \propto w_{k-1}^{\TR (i)}\ \frac{ p(\bz^{\TR}_{k} \mid \bx^{\TR (i)}_{k})\ p(\bet^{\LR}_{k} \mid \bx^{\TR (i)}_{k})\ p(\bx^{\TR (i)}_{k} \mid \bx^{\TR (i)}_{k-1})}{q(\bx^{\TR (i)}_{k} \mid \bx^{\TR (i)}_{k-1}, \bz^{\TR}_{1:k}, \bet^{\LR}_{2:k})} .
\ee